%% file: ms.tex
\pdfoutput=1
\documentclass[preprint,11pt,a4paper]{elsarticle}

\usepackage[T1]{fontenc}
\usepackage[utf8]{inputenc}
\usepackage[a4paper,margin=2.2cm]{geometry}
\usepackage{amsmath,amssymb}
\usepackage{booktabs,longtable,array,tabularx}
\usepackage{graphicx}
\usepackage{microtype}
\usepackage{xcolor}
\usepackage{setspace}
\usepackage{lineno}
\usepackage{tikz}
\usepackage{hyperref}
\hypersetup{colorlinks=true,linkcolor=blue,citecolor=blue,urlcolor=blue}
\usetikzlibrary{arrows.meta,positioning,shadows.blur,fit}

\graphicspath{{./figures/}}
\journal{Simulation Modelling Practice and Theory}

\begin{document}

\begin{frontmatter}

\title{Data-driven regional traffic simulation from heterogeneous mobility sources}

\author[ifisc]{Federico Bellisardi\corref{cor1}}
\ead{fbellisardi@ifisc.uib-csic.es}
\cortext[cor1]{Corresponding author.}
\author[mindicity]{Chiara Mizzi}
\author[mindicity]{Alessandro Fabbri}
\author[telecom]{Davide Micheli}
\author[telecom]{Aldo Vannelli}
\author[unibo,infn]{Armando Bazzani}

\affiliation[ifisc]{organization={Instituto de F\'{\i}sica Interdisciplinar y Sistemas Complejos IFISC (CSIC-UIB)},
  addressline={Campus UIB},
  postcode={07122},
  city={Palma de Mallorca},
  country={Spain}}
\affiliation[mindicity]{
  organization={Mindicity, fabbricadigitale S.r.l.},
  city={Casalmaggiore (CR)},
  country={Italy}
}
\affiliation[telecom]{
  organization={TIM -- Telecom Italia S.p.A.},
  city={Rome},
  country={Italy}
}
\affiliation[unibo]{organization={Laboratory of City Science, Physics and Astronomy Department ``Augusto Righi'', University of Bologna},
  addressline={Viale Carlo Berti Pichat 6/2},
  postcode={40127},
  city={Bologna},
  country={Italy}}
\affiliation[infn]{organization={INFN, Sezione di Bologna},
  addressline={Viale Carlo Berti Pichat 6/2},
  postcode={40127},
  city={Bologna},
  country={Italy}}

\begin{abstract}
Regional traffic simulators increasingly combine mobile-phone, road-sensor and GPS data, but what each source contributes is rarely clear. We present an agent-based simulator of daily traffic in Emilia-Romagna, Italy: phone data give trips and their timing, half of the road sensors set the traffic level, all sensors correct it locally, and GPS traces guide route choice. We measure what each source adds against predictors without simulation, on sensors kept out of parameter calibration in winter 2020 and partly shared with it in summer 2021; as these sensors informed some modelling choices, the evaluation is not fully independent. Total traffic is within 1\% of observations in winter and 10\% in summer, and the simulator ranks roads better than a road-type predictor (Spearman 0.67--0.68 against 0.41--0.48, winter). GPS improves this ranking, sensors improve local estimates, and timing comes mostly from the demand. Single-road errors remain similar to those of simple predictors, and the test of specific origin--destination pairings was underpowered and inconclusive. These findings provide a practical framework for assessing the contribution and limitations of heterogeneous data sources in regional traffic simulation.
\end{abstract}

\begin{keyword}
Agent-based simulation \sep Simulation validation \sep Traffic reconstruction \sep Mobile phone data \sep Loop detector \sep GPS probe data
\end{keyword}

\end{frontmatter}

\section{Introduction}
\label{sec:intro}

Transport planning \cite{toole2015}, the study of congestion \cite{colak2016} and the assessment of emissions \cite{grujic2022} all need to know how much traffic each road carries, hour by hour. No single data source provides this. Mobile phones show where people travel \cite{gonzalez2008,calabrese2011} but not which roads they use, and phone samples have their own biases \cite{wesolowski2013}. Loop detectors count every vehicle, but only at a few points, and where they are placed limits what we can learn about the rest of the network \cite{yang1998,castillo2008,gentili2012,castillo2015}. GPS-equipped vehicles cover many roads, but they are a small sample of traffic, and the chance of sampling a vehicle changes between roads and hours \cite{herrera2010,jenelius2013,patire2015}. A regional model therefore has to combine these sources \cite{elfaouzi2011,zheng2014,chen2016,anda2017}. This raises a simple question: what does each of them actually add?

Each source has been studied for a long time. Phone and GPS data have shown statistical regularities in how individuals move \cite{bazzani2010,gallotti2016,mizzi2018,mizzi2022,mizzi2023}. Phone records have also been used to estimate origin--destination (OD) matrices \cite{iqbal2014,alexander2015}, and these matrices have been assigned to road networks \cite{wardrop1952} to study road use and congestion \cite{wang2012,toole2015,colak2016}. Detector counts have long been used to estimate OD matrices \cite{cascetta1984,yang1992,antoniou2016} and to calibrate agent-based simulations \cite{flotterod2011}. GPS probes and fixed sensors are used to estimate the state of traffic, mostly on motorways and urban corridors \cite{wang2005,work2010,bachmann2013,seo2015,nantes2016,seo2017}.

More recent work combines the sources. Telecom data and vehicle flows have been used together to predict traffic where there is no camera \cite{lin2024}, traffic has been estimated at places without sensors \cite{lei2024}, and a few fixed sensors have been scaled up to describe whole urban networks \cite{maiti2025}. Closer to our work, agent-based simulations now take their demand from phone data. In Barcelona, activity diaries built from phone records feed a MATSim model \cite{bassolas2019}. In Novi Sad, telecom data are combined with population and GIS data, and traffic is calibrated and validated with vehicle counters \cite{grujic2022}. Aggregated phone data have also been used to build a model of Brussels \cite{li2024}. A recent review of more than 60 such case studies shows how common these data-driven, city-wide simulations have become \cite{bochenina2025}. Combining the sources is therefore no longer new, but doing so does not by itself show what each of them adds.

Three things make this hard to see. First, independent data are scarce. The simulation literature separates verification, whether a model is implemented as intended, from validation, whether it is accurate enough for its purpose, and asks that validation use data not used to build the model \cite{kleijnen1995,sargent2013}; the same holds for traffic simulation \cite{toledo2004,hollander2008}. Yet demand is often rescaled to match the counts, and agreement at those same counts is then reported. The closest studies show how few options there can be: \c{C}olak et al.\ report that no single validation source existed for their cities \cite{colak2016}, Toole et al.\ compare their estimates with travel surveys \cite{toole2015}, and Bassolas et al.\ keep traffic counts out of the calibration but have them only at some main entrances of the city \cite{bassolas2019}. Second, the sources are usually tested together. The result then shows whether the complete model works, not which source made it work, even though uncertainty in demand affects every stage of a model \cite{zhao2002,rasouli2012}. Third, a model needs a benchmark. Simple historical profiles predict the daily pattern of traffic very well \cite{lippi2013,vlahogianni2014}, so good timing means little unless a model does better than they do.

In this paper we build and test such a simulator for a whole region, Emilia-Romagna, Italy, with hourly traffic on the regional road network. We give each source a defined role in demand estimation, route choice or local correction, and check each role with data that did not set its values (Table~\ref{tab:roles}). Combining the sources is not new; what we add is an evaluation of what each of them contributes to the reconstruction and of whether the simulation does better than simpler predictors. The contributions are methodological as much as regional:
\begin{itemize}
\item an assessment protocol that keeps apart five objectives that are often reported together: the aggregate level of traffic, the ranking of roads, the daily timing, the estimate at roads without detectors, and the contribution of each source;
\item a measure of the marginal information of each source, the change on held-out data when that source is added, changed or removed, with the parameters that act through it held fixed;
\item a comparison of every output with predictors that use no simulation, so that agreement with the data is not mistaken for skill;
\item an account of what the evidence does not establish, including the error at single roads and the value of specific origin--destination pairings.
\end{itemize}
Half of the detectors scale the demand, with one factor per day type and one per province. Route choice uses GPS measured in one province and transferred to the rest of the region by road class. We compare timing on the GPS-matched roads of Forl\`i--Cesena, volumes at held-out detectors in two seasons, and a local correction by leaving out each detector in turn, always against predictors that use no simulation. Section~\ref{sec:data} describes the data, Section~\ref{sec:methods} the model and the tests, Section~\ref{sec:results} the results in the order in which the model is built, and Section~\ref{sec:discussion} what they mean.

\begin{table}[htbp]
\centering
\caption{Role of each data source. Each row gives one data set, the part of the model it sets and the comparison it serves. Each step is compared with data that did not set its values; notes~$a$--$c$ give the cases in which comparison data informed a modelling choice. The detectors of 2021 include 60 of the calibration half.}
\label{tab:roles}
\begin{tabular}{@{}p{3.6cm}p{5.4cm}p{6.0cm}@{}}
\toprule
Data & Enters the model in & Used for comparison in \\
\midrule
Mobile-phone OD (Jan--Feb 2020) & where and when trips occur & -- \\
GPS traces, Forl\`i--Cesena & route choice (class speeds and usage) & -- \\
GPS link counts, Forl\`i--Cesena, two campaigns & choice of the form of the route cost$^{c}$ & hourly profiles on the matched links \\
Detectors, half of February 2020 & demand scale $k$, province factors, trip radius, departure shift, cost weights, depth$^{a}$ & -- \\
Detectors, other half of February 2020 & local correction; adoption of province factors$^{b}$ & volume level and ranking (2020) \\
Detectors, June--July 2021 & local correction; adoption of province factors$^{b}$ & volume level and ranking, later season \\
Detector left out of the correction & -- & leave-one-detector-out test \\
\bottomrule
\end{tabular}
\par\vspace{2pt}{\footnotesize $^{a}$The depth of the correction was selected by cross-validation over all detectors of 2020; 2021 played no part in that selection. $^{b}$These detectors decided once whether the factors, set on the calibration half, were kept; they set no value. $^{c}$Daily GPS counts of February 2020 selected the form of the cost; the comparison uses hourly shares, which route choice hardly changes.}
\end{table}

\begin{figure*}[htbp]
\centering
\begin{tikzpicture}[font=\footnotesize, >={Latex[length=2mm]},
  inp/.style={draw, rounded corners=2pt, fill=blue!7, align=center, minimum height=8mm, text width=3.3cm},
  proc/.style={draw, fill=orange!15, align=center, minimum height=8mm, text width=3.8cm},
  chk/.style={draw, dashed, fill=gray!10, align=center, minimum height=8mm, text width=3.6cm}]
\node[inp] (od) at (0,3.3) {Mobile-phone OD tables};
\node[inp] (cal) at (0,2.2) {Detectors, calibration half};
\node[inp] (gps) at (0,1.1) {GPS traces, one province};
\node[inp] (net) at (0,0) {OpenStreetMap road network};
\node[inp] (det) at (0,-1.1) {Detectors, all others in turn};
\node[proc] (dem) at (5,3.3) {Demand by hour and class};
\node[proc] (par) at (5,2.2) {Scale $k$, province factors, radius, departure shift};
\node[proc] (rou) at (5,1.1) {Class speeds, usage and cost weights};
\node[proc] (sim) at (5,0) {\texttt{minimocas}: hourly traffic on all links};
\node[proc] (cor) at (5,-1.1) {Local correction};
\node[chk] (t1) at (10.8,1.1) {Volumes: test half 2020, detectors 2021};
\node[chk] (t2) at (10.8,0) {Timing: GPS link counts, one province};
\node[chk] (t3) at (10.8,-1.1) {Error at left-out detectors};
\foreach \a/\b in {od/dem,cal/par,gps/rou,net/sim,det/cor,dem/par,par/rou,rou/sim,sim/cor} \draw[->] (\a) -- (\b);
\draw[->, dashed] (sim) -- (t2);
\draw[->, dashed] (8,0) |- (t1);
\draw[->, dashed] (cor) -- (t3);
\end{tikzpicture}
\caption{\textbf{Pipeline of the simulator.} Read from left to right: each input feeds one step of the model, and the steps together produce hourly traffic on all links, which is then compared with data on the right. Blue: input data. Orange: steps of the model. Grey, dashed: data used to assess the output, which do not set the values of the steps they assess.}
\label{fig:pipeline}
\end{figure*}
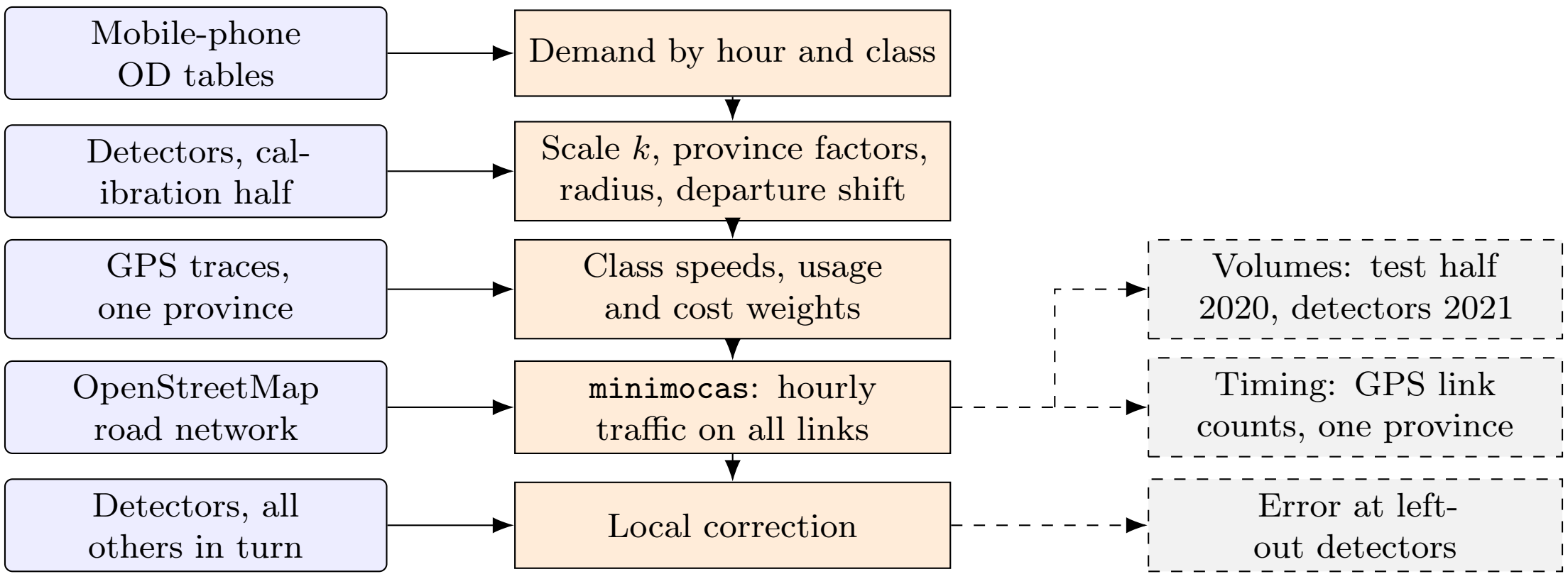

Figure~\ref{fig:pipeline} shows how the data enter the simulator and which data are used to assess each output.

\section{Data}
\label{sec:data}

\begin{figure*}[htbp]
\centering
\includegraphics[width=\textwidth]{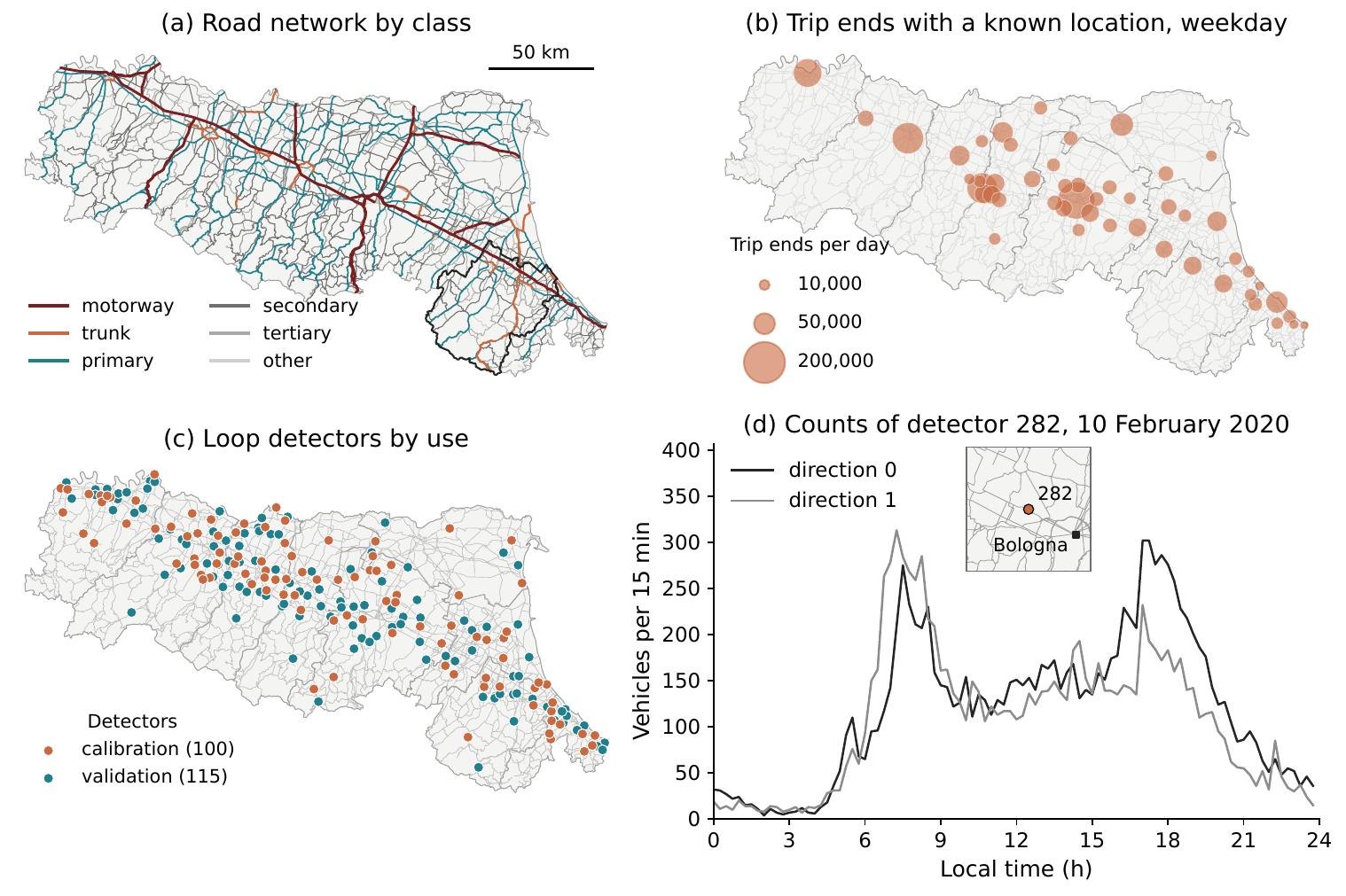}
\caption{\textbf{The region and the data that enter the model.} \textbf{(a)} Road network coloured by OpenStreetMap road class; the province of Forl\`i--Cesena, where GPS traces inform route choice, is outlined. \textbf{(b)} Daily trip ends with a known location in the weekday demand, by municipality: a larger circle means more trips, and the demand concentrates along the central axis of the region and on the coast. \textbf{(c)} Loop detectors of the regional MTS system placed on the network, by their use for daily volumes: the calibration half of February 2020 (100; 60 of them also enter the 2021 tests) and validation only (115: the 97 of the test half and 18 used only in 2021). The other 77 detectors of the network have no complete counts in either campaign and are not shown. \textbf{(d)} Counts of detector 282 near Bologna on 10 February 2020, 15-min resolution, local time. Road network \copyright{} OpenStreetMap contributors.}
\label{fig:inputs}
\end{figure*}

The road network comes from OpenStreetMap \cite{osm}, whose road coverage has been estimated to exceed 95\% in most European countries \cite{barrington2017}. It has 3,925 links and 2,548 nodes, with their length, direction and turn restrictions, and each link has an OpenStreetMap road class (Fig.~\ref{fig:inputs}a). 

For the demand, the operator TIM provided hourly activity between municipalities for January--March 2020. We used 13 January--23 February, which excludes the holidays of early January and the restrictions for COVID-19 that began in the region on 23 February, and averaged it separately for each day of the week: each simulated day uses the mean of its own day of the week, so a Tuesday is simulated with the mean of the six Tuesdays of the period. The model keeps the relations among the 50 most active municipalities and between them and the provinces of neighbouring regions. When one end of a trip, or both, is outside this set, the model draws the missing end at random. We left out trips within each of the 50 municipalities, because adding them did not reduce the spread of the errors on the calibration detectors (Supplementary Section~S2). Trips within the other municipalities stay in the demand, with both ends drawn at random.

So only part of the recorded activity enters the model as a relation. We give the figures for the mean Monday; those for the mean Saturday are similar (Supplementary Section~S8). The Monday table holds 3.15 million trips per day, in three parts. 18\% have the code 000000000 at one or both ends; this code identifies no place, and we remove these trips. 43\% are within a municipality: those within the 50 municipalities (38\% of the table) are left out, and those within the other municipalities (5\%) enter with both ends drawn at random. The remaining 39\% are between municipalities: 28\% of them keep their relation, and 72\% have a recorded relation that the model does not use and enter with their end in the selected set, if they have one, and the other end drawn at random. The result is 1.99 million agents on a Monday and 1.61 million on a Saturday, after scaling by $k$ and the province factors. Of the Monday agents, 28\% have a known origin and destination and 72\% have at least one random end (Supplementary Fig.~\ref{fig:demand} and Section~S1). We also corrected a double counting of the trips with neighbouring provinces in the configuration generator we inherited (Supplementary Section~S2). Since the demand depends only on the day of the week, the fifteen days of the two campaigns need one simulation for each day of the week, plus a second Monday that checks that a run is reproducible. The 2021 days therefore use demand measured in early 2020. The hour labels in the tables refer to local time at the start of the hour (Supplementary Section~S7).

The regional MTS system counts vehicles every 15~min in both directions, and 292 of its detectors lie on links of the network (Fig.~\ref{fig:inputs}c,d). A detector is used on a given day only if both of its directions are complete. This filter favours busy roads, but relaxing it hardly changes the errors (Supplementary Section~S7).

Viasat provided hourly probe counts per matched link for two campaigns, 10--17 February 2020 (six weekdays and a weekend) and 28 June--4 July 2021 (five weekdays and a weekend), and the raw traces of private vehicles in the province of Forl\`i--Cesena. We use the link counts of that province only (Supplementary Section~S5). Probes were not used to generate or scale the demand, or in the detector correction.

Phone data were provided as anonymized counts aggregated in space and time, and we accessed no individual records. GPS data were provided as pseudonymized vehicle records and used only to compute link counts, speeds and road-class aggregates.

\section{Methods}
\label{sec:methods}

\subsection{Simulation and route choice}
The simulator moves agents in steps of 10~s, each at the speed of its link. This speed falls as the link becomes denser, down to a critical speed. Agents with a known origin and destination follow best paths computed with Dijkstra's algorithm \cite{dijkstra1959}, which respect turn restrictions. Agents with a random end instead choose the next link at each node, with a probability proportional to a transition weight, and travel for a time drawn between 900 and 5,600~s. Knowing their destination determines where the trip starts, but not the route they take. For each hour and direction the simulator records how many distinct agents used each link, while virtual counters at the detector locations record crossings. Each run starts empty at midnight, and a given random seed always gives the same run. Traffic is not assigned to an equilibrium \cite{wardrop1952}, and the model does not reproduce the congestion phenomena of microscopic traffic-flow models \cite{treiber2013,lopez2018} or the queue dynamics of agent-based transport simulators such as MATSim \cite{horni2016}.

To inform route choice, we matched GPS fixes of private vehicles in Forl\`i--Cesena (February 2020) to the road network. For each road class we took the median off-peak speed and the median daily number of probes per link (Supplementary Section~S3), and applied both to all links of that class in the region. The speeds seem stable over time: on the same roads, those of June 2021 have a rank correlation of 0.98 with those of February 2020. The probe numbers set the transition weights of agents with a random end. They also enter the best-path cost of link $i$,
\begin{equation}
w_i = L_i\left[\alpha + \alpha_s\,\frac{v_0}{v_i} + (1-\alpha-\alpha_s)\,\frac{1/c_i - 1/c_{\max}}{1/c_{\min} - 1/c_{\max}}\right],
\label{eq:route_cost}
\end{equation}
where $L_i$ is its length, $v_i$ its speed, $v_0=20$~km/h a reference speed of the simulator that makes the speed term dimensionless (only the product $\alpha_s v_0$ enters, so $v_0$ only sets the scale of $\alpha_s$), and $c_i$ the usage weight of its class, so that busier road classes cost less. This cost describes an empirical preference by road class, and it is not a behavioural model. We chose the usage-weighted form among shortest-distance, travel-time and mixed alternatives because of its rank correlation between simulated daily volumes and GPS counts in Forl\`i--Cesena, without using detectors. The weights $\alpha=0.5$ and $\alpha_s=0.25$ come from a discrete grid compared on the calibration detectors; they are chosen values, not estimates (Supplementary Section~S3).

\subsection{Implementation, verification and computational cost}
\label{sec:implementation}
The simulator, \texttt{minimocas}, is written in C++14. Its inputs are the road graph with turn restrictions, the speed and usage weight of each road class, the trip sources with their creation rate in each 15-min interval, and the detector positions. Before the run it computes a best-path matrix for each class of agents with Dijkstra's algorithm, in parallel over nodes with OpenMP. Agents are then moved one after the other in each 10-s step, without parallelism, from 00:00 to 23:45 (8,550 steps).

We verified the implementation in three ways. A given random seed always gives the same run: the second simulated Monday of the 2020 campaign is identical to the first. After a defect in a diagnostic output was corrected, the engine reproduced the link flows of the final Monday byte for byte. The configuration generator, rewritten because the original was not available, writes the configurations only after it has reproduced the creation rates of the inherited ones for the same period. Between four random seeds, the volume ratio and the Spearman correlation on held-out detectors change by at most 0.002 and the dispersion by at most 0.006 (Supplementary Section~S8), well below the differences we interpret.

On a desktop computer (AMD Ryzen 3 3200G, four cores), one simulated Monday with 1.99 million agents took 7.1~min of wall-clock time, 6.5~min of which to move the agents, with a peak memory of 2.5~GB; it used about one core, with two OpenMP threads for the best paths. Its link flows and counter crossings were identical, byte for byte, to those of the final run used in this paper. Simulated days are independent, so the days of a campaign run in parallel as separate processes. We did not measure how the cost grows with the size of the network or the number of agents.

\subsection{Demand scaling and local correction}
Detectors enter the model in two ways. Half of them set the overall level of the demand, with one factor for each day type and one for each province. The counts of each day then correct the simulation locally, near each detector.

For the first use, we split the detectors with complete data in February 2020 into a calibration half (100 detectors) and a test half (97), with the same share of each province in both. Only the calibration half sets values; the test half and the 2021 campaign are kept for the tests.

The OD tables describe one operator's customers, not all traffic, so we multiply the number of agents by a factor $k$. This factor combines coverage, vehicle occupancy and the effect of random trip ends, and has no physical meaning of its own. In particular, it trades off against trip length: simulated agents travel about 40~km on a Monday, while the distances between municipalities in the tables average 12~km in a straight line (Supplementary Section~S8). We compare a Monday simulation with the mean weekday counts and a Saturday simulation with the mean weekend counts, and for each we choose the $k$ that minimizes the mean squared $\log_{10}$(simulated/observed) of daily volumes, on a grid from 0.9 to 1.8 (Supplementary Section~S2).

The same loss chose the radius within which trip ends are drawn, among 15, 20, 25 and 30~km. Random ends lie within 25~km of the known end, and the ends of a known relation within 25~km of its two municipalities, weighted by population, so a relation fixes the area of a trip and not its exact ends. Origins are also spread around their source point. Departures are 45~min earlier than the OD label: this reduced the hourly error on the Monday but not on the Saturday, and we apply it to both (Supplementary Section~S2).

Even with the level set, the error still varied strongly from road to road. We measure this with the \emph{dispersion}, the standard deviation of $\log_{10}$(simulated/observed) of daily volumes across detectors. On the calibration half, 38\% of its variance was between provinces (Section~\ref{sec:single_roads}), so we tested one more factor per province. Each trip source with a known location is multiplied by the geometric mean of the factors of its ends inside the region. Sources without a location and ends outside the region keep a factor of one, and $k$ does not change. The factor of a province starts from one and is multiplied by $10^{-m}$, where $m$ is the median $\log_{10}$(simulated/observed) of its calibration detectors over both day types. We do this for two rounds. We fixed the rule for adopting the factors before the runs: the 95\% paired bootstrap interval of the change in dispersion had to be below zero on the test half and, only if it was, on the detectors of 2021. The factors add nine parameters, and their values come from the calibration half only.

For the second use, where there is a detector, we can replace the simulated count with the observed one and spread the difference to nearby roads. Let $u_i(h)$ be the simulated count on the directed link of detector $i$ in hour $h$ and $o_i(h)$ the observed count in that direction; detectors on the same directed link are averaged. The increment at the detector link is
\begin{equation}
\delta_i(h)=\begin{cases} o_i(h)-u_i(h) & \text{if } u_i(h)\ge 1,\\ 0 & \text{otherwise,}\end{cases}
\label{eq:coil_residual}
\end{equation}
so the corrected count $\widetilde u_i(h)=u_i(h)+\delta_i(h)$ equals the observed one wherever the simulation has traffic at the detector. From there the increment spreads downstream and upstream. Each adjacent link receives the increment of the previous link, multiplied by its own simulated share of the flows leaving (downstream) or entering (upstream) the common node, $\delta_{(n_2,n)}=\delta_{(n_1,n_2)}\,u_{(n_2,n)}/\sum_m u_{(n_2,m)}$. The spreading does not enter other detector links. It stops when the increment falls below one vehicle, after three links along a path, at a forbidden turn or at a closed link. Any negative count is set to zero.

\subsection{Evaluation metrics and tests}
\label{sec:assessment_methods}
To assess timing, we compare on each GPS-matched link the hourly shares $p_i(h)$ and $q_i(h)$ of the reconstructed and GPS counts over the campaign days, both directions together,
\begin{equation}
\mathrm{MAE}_i=\frac{100}{24}\sum_{h=0}^{23}|p_i(h)-q_i(h)|,
\label{eq:profile_error}
\end{equation}
in percentage points (pp), against three benchmarks: the OD profile of the demand, the mean profile of the detectors and a flat profile.

Daily volumes are compared at detectors with complete data, both directions, by the ratio of total simulated to total observed volume, the Spearman rank correlation, the median absolute $\log_{10}$ ratio and the share of detectors within a factor of two, with intervals from a paired bootstrap over detectors. The predictors without simulation take the level from the median volume of other detectors of the same road class and the profile from their mean profile. For the test half they are built on the calibration half, and for the leave-one-out test on the other detectors of the same day.

In the leave-one-out test, we leave out all detectors of a link in turn and compute the error $\sum|\widetilde u(h)-o(h)|/\sum o(h)$ of the corrected reconstruction on that link, over hours, directions and days. The depth of the correction, the number of links it spreads along, is the smallest whose error reduction in 5-fold cross-validation over 2020 is within one bootstrap standard error of the best (Supplementary Section~S7).

The final tests use the raw simulation on the test half of 2020 and on all detectors of 2021, and each supports only the property it tests. For the level, the volume ratio is equivalent to one within 0.80--1.25 if its 90\% bootstrap intervals over detectors and over provinces both lie within that margin, as in two one-sided tests; since we chose the margin, we also report the smallest one that passes. For the ranking of roads, we compare the Spearman correlation with permutations of the simulated volumes, among all detectors and only within road class; the second test looks only at what road class alone does not explain. For timing, the hourly-share error on the links of Forl\`i--Cesena should be below that of the flat profile (one-sided Wilcoxon signed-rank test). The p values of the ranking and timing tests are Holm-adjusted over the six tests, three in each campaign.

These data are held out from parameter calibration but not blind, and the evaluation subsets differ in how far they are separated from model development. (i) The calibration half sets every numerical value (Table~\ref{tab:roles}). (ii) The test half of 2020 sets no value, but it decided whether the province factors were adopted, and the removal of the code 000000000 and the travel-time term in the cost were decided after earlier versions had been compared on it. (iii) The 141 detectors of 2021 played the same part in these decisions, and 60 of them belong to the calibration half, so we also report the 81 others. (iv) In the leave-one-out test a detector never enters its own correction, but the 2020 set includes the calibration detectors and selected the depth of the correction. No subset is therefore fully independent of model development; the 81 detectors of 2021 outside the calibration half come closest, as they depend on it only through these decisions. The results are held out rather than pre-registered.

The experiments on the OD relations followed protocols written before their runs, are exploratory and do not change the model; they are described with their results in Supplementary Section~S9.

\section{Results}
\label{sec:results}

\subsection{Level and timing of traffic}
\begin{figure*}[htbp]
\centering
\includegraphics[width=\textwidth]{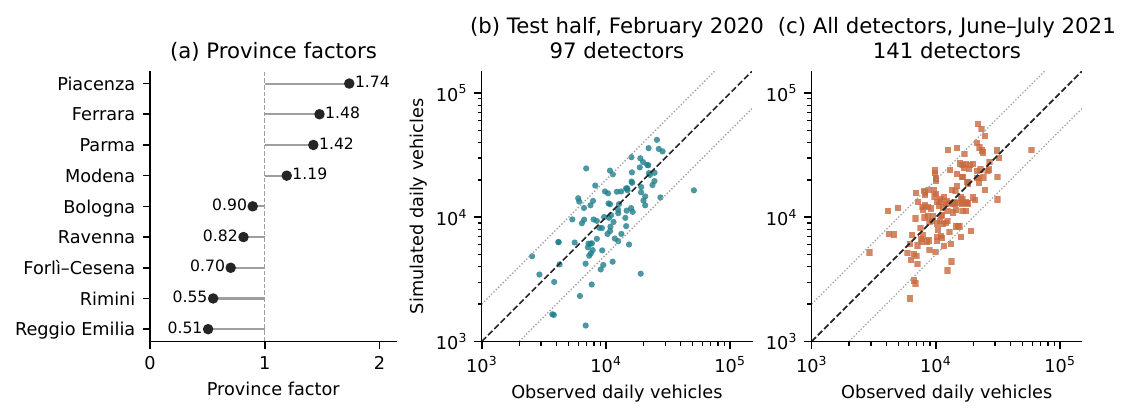}
\caption{\textbf{The simulator reaches the overall level of traffic.} \textbf{(a)} Factor applied to the demand of each province; values above one (dashed line) raise the demand. The factors were set on the calibration half of the detectors. \textbf{(b,c)} Daily volume at each detector, simulated against observed, raw simulation before any local correction, on logarithmic axes. A point on the dashed line would be a perfect estimate, and points between the dotted lines are within a factor of two. A cloud centred on the dashed line means that the level is right overall, and its width shows the error at single roads. Panel (b) uses detectors kept out of parameter calibration; panel (c) uses a later season and includes 60 calibration detectors. One detector in each panel has no simulated traffic and is not drawn.}
\label{fig:level}
\end{figure*}

We first ask whether the simulator puts the right amount of traffic on the network. Scaling the demand by day type, with $k=1.5$ on weekdays and $k=1.3$ on weekends, brings simulated and observed volume on the calibration detectors within 7\% and 10\%, and within 1\% and 4\% with the province factors. A single common factor would leave weekends 18\% too high, and the choice is robust to subsets of detectors (Supplementary Fig.~\ref{fig:k_choice} and Section~S8). Regional differences remain after this step, and the province factors correct them. Figure~\ref{fig:level}a shows that the correction is far from uniform: demand is raised in the north-west, in Modena and in Ferrara and lowered along the east and in Reggio Emilia. We treat these factors as empirical, not as measured differences in phone coverage or car use.

Figures~\ref{fig:level}b and \ref{fig:level}c show the outcome at detectors that did not set these values, or set them only in part in 2021. In both campaigns the cloud of points follows the diagonal: the overall level is right and busy roads tend to be simulated as busy, while the width of the cloud shows the error at single roads. The province factors met the rule for adoption in both data sets: dispersion falls from 0.500 to 0.489 on the test half and from 0.415 to 0.399 in 2021, and the gain is clearest at the typical detector, where the median error falls by about a fifth, from 0.180 to 0.148 and from 0.177 to 0.142 (Supplementary Table~\ref{tab:province_effect}). These are the data on which the adoption was decided, so this comparison supports the choice but is not a separate test of it. Factors estimated on 25 or 50 calibration detectors mostly give a dispersion close to that of the 100 used here, although the value for a single province varies strongly between subsets, so the factors should be read as effective corrections and not as precise estimates (Supplementary Section~S8). All later results use the model with the factors.

\begin{figure*}[htbp]
\centering
\includegraphics[width=\textwidth]{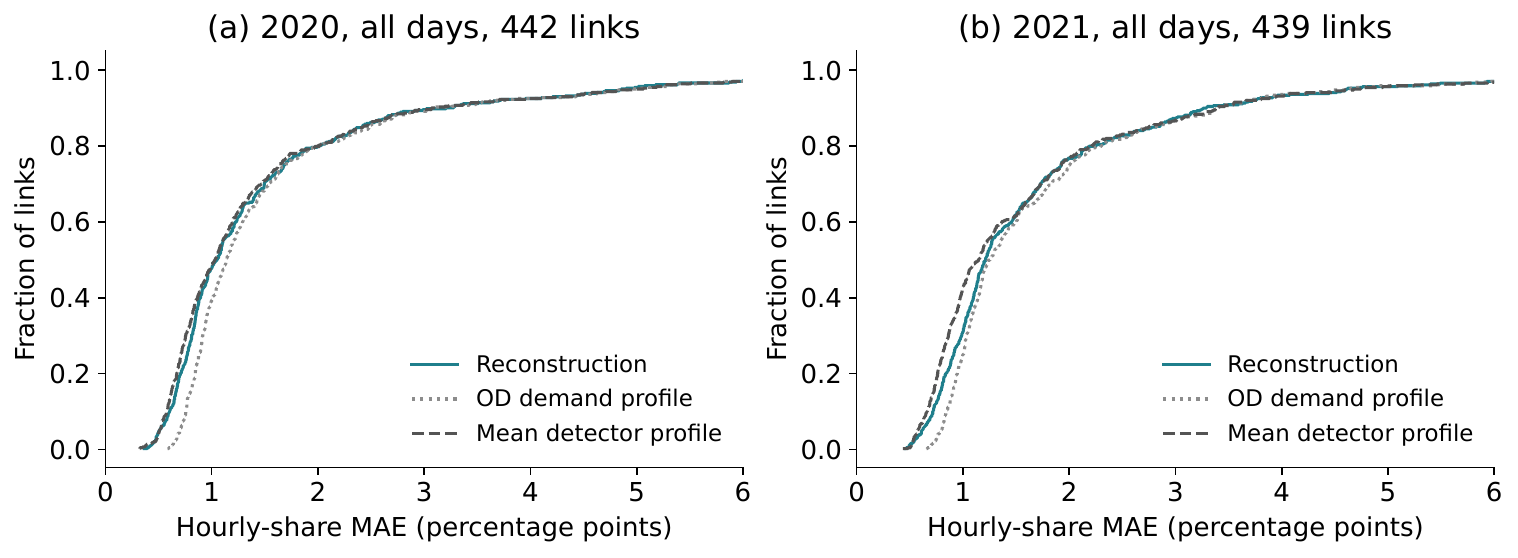}
\caption{\textbf{The timing of traffic follows GPS on nearly all links.} For each GPS-matched link we compute the error of its hourly shares, the mean absolute difference between the share of daily traffic in each hour in the estimate and in GPS. Each curve shows the fraction of links whose error is below the value on the horizontal axis, so a curve further up and to the left is better. Reconstruction: simulation after local correction. \textbf{(a,b)} All days, against the OD demand profile and the mean detector profile, in 2020 and 2021. The curves are close, and a minority of links has large errors for all three. Curves continue beyond the 6~pp limit of the axis.}
\label{fig:timing_cdf}
\end{figure*}

With the level set, we turn to when traffic occurs, using GPS as a view of the daily rhythm on the matched roads of Forl\`i--Cesena. Figure~\ref{fig:timing_cdf}a,b compares the simulation with two simple alternatives: the hourly profile of the demand itself and the average profile of the detectors. All three are far better than a flat profile, and the curves are close: most of the daily rhythm is already in the demand. The simulation improves on the demand profile, and after correction it stays slightly behind the detector average in both campaigns (median errors 1.08 and 1.21~pp raw, 1.05 and 1.21~pp corrected, 1.03 and 1.15~pp for the detector average; Supplementary Table~\ref{tab:gps_split}). Weekend timing is harder than weekday timing for both the simulation and the detector average (Supplementary Fig.~\ref{fig:daytypes}). GPS traces of this province also set the route weights, and its daily GPS counts chose the form of the route cost, so this comparison is not fully independent of route choice, although route choice hardly changes the profiles (Supplementary Section~S3).

\begin{figure*}[htbp]
\centering
\includegraphics[width=\textwidth]{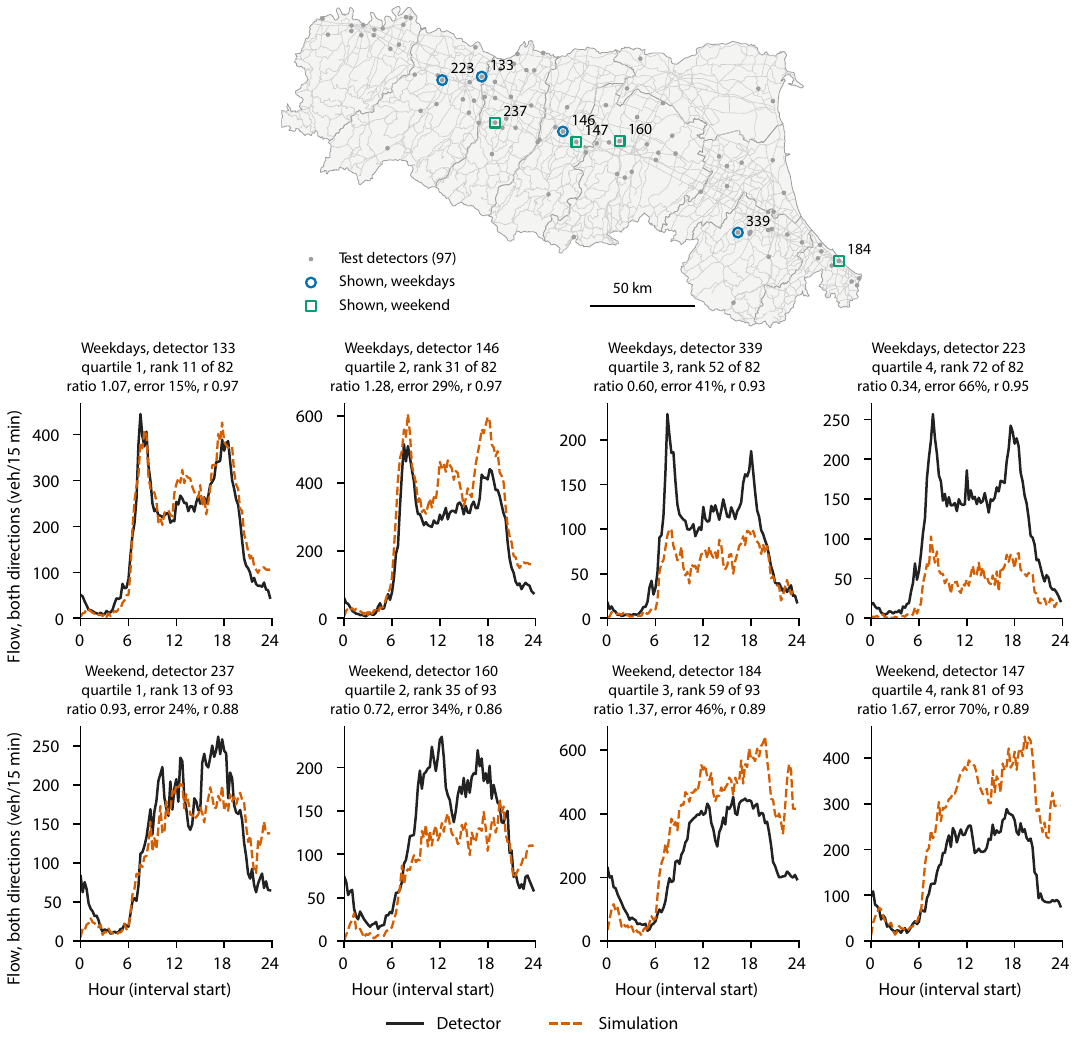}
\caption{\textbf{Daily profiles at test detectors across the range of the error.} Top: road graph, test detectors and the detectors shown below, by day type. Below: one test detector from the middle of each quartile of the 15-min error, from the lowest (left) to the highest (right). Mean observed flow (solid) and raw simulation before any correction (dashed), both directions, February 2020, 15-min resolution, for weekdays (top) and weekends (bottom). The title of each panel gives the quartile and rank among test detectors, the ratio of simulated to observed volume, the 15-min error and the correlation of the two curves. The median detector has a 15-min error of 33\% on weekdays and 38\% on weekends.}
\label{fig:detector_profiles}
\end{figure*}

Detectors count all vehicles, so they let us look at the actual flow through the day. Figure~\ref{fig:detector_profiles} shows four test detectors for each day type, one from the middle of each quartile of the 15-min error, and where they lie in the region.
On weekdays the simulation follows both peaks and the midday plateau at all four detectors, also where the level is wrong (profile correlations 0.93--0.97), so the error comes mostly from the level. On weekends it follows the general shape less closely, starting too low at night and running too high late in the evening. Table~\ref{tab:baselines_detectors} gives the picture over all test detectors. The hourly error is about 40\%, and it would fall by about two thirds on weekdays if each detector were given its observed daily level, so at a single road the difference is mostly how much traffic there is, not when. All detectors share a similar daily pattern, and the plain average of the calibration detectors has a higher profile correlation with each test detector than the simulation has. The simulation, however, ranks the roads clearly better than a predictor based on road class.

\begin{table}[htbp]
\centering
\caption{Raw simulation against a predictor without simulation, test detectors of February 2020. A volume ratio close to one, lower errors and higher correlations are better. The predictor, built on the calibration detectors, is the median volume of the same road class times the mean profile. The province factors were also set on those detectors. Volumes and errors come from the virtual counters, both directions; these counters are about 8\% below the distinct-agent counts on which $k$ is set. Intervals: 95\% bootstrap of the difference, simulation minus predictor. Peaks, simulated against observed: weekdays 8:00 and 18:00 against 7:30 and 17:30; weekend evening 19:15 against 17:30.}
\label{tab:baselines_detectors}
\begin{tabular}{@{}lrrlrrl@{}}
\toprule
 & \multicolumn{3}{c}{Weekdays (82 detectors)} & \multicolumn{3}{c}{Weekends (93 detectors)} \\
Measure & Sim. & Pred. & Interval & Sim. & Pred. & Interval \\
\midrule
Volume ratio & 0.93 & 0.92 & & 0.93 & 0.88 & \\
Median $|\log_{10}|$ ratio, daily volume & 0.151 & 0.137 & $[-0.05, +0.05]$ & 0.145 & 0.157 & $[-0.06, +0.03]$ \\
Within a factor of two (\%) & 83 & 89 & & 81 & 88 & \\
Spearman, daily volume & 0.67 & 0.41 & $[0.06, 0.46]$ & 0.68 & 0.48 & $[0.02, 0.36]$ \\
Hourly error (\%) & 37.9 & 40.1 & $[-11.1, +6.6]$ & 43.0 & 41.1 & $[-6.0, +9.5]$ \\
Hourly error with known level (\%) & 12.9 & 7.2 & & 22.4 & 8.2 & \\
Median profile correlation & 0.96 & 0.99 & $[-0.04, -0.03]$ & 0.86 & 0.98 & $[-0.12, -0.10]$ \\
\bottomrule
\end{tabular}
\end{table}

\subsection{Daily volumes and local correction}
Since most of the difference at detectors is about level, we now look at daily volumes (Table~\ref{tab:volumes}, top). The results of the two campaigns are similar: the total volume is close to the observed one, the ranking of roads is good and more than four detectors in five are within a factor of two. The level in 2021 is reached with demand measured in early 2020 and a scale fixed on winter counts, so it should not be read as a test of demand in that season. Without the 60 calibration detectors, the 81 remaining detectors of 2021 give a volume ratio of 1.11 [0.99, 1.24], a Spearman correlation of 0.65 [0.50, 0.76] and 81\% [73, 90] within a factor of two. Road class adds a further pattern (Supplementary Fig.~\ref{fig:distributions}): the simulation underestimates secondary roads (median ratio 0.78 in both campaigns) and tends to overestimate primary roads slightly (1.05 and 1.12), while trunk and tertiary roads have too few detectors for a clear value. No detector lies on a motorway, so the level on motorways, 709~km of the network, is not tested.

\begin{figure*}[htbp]
\centering
\includegraphics[width=\textwidth]{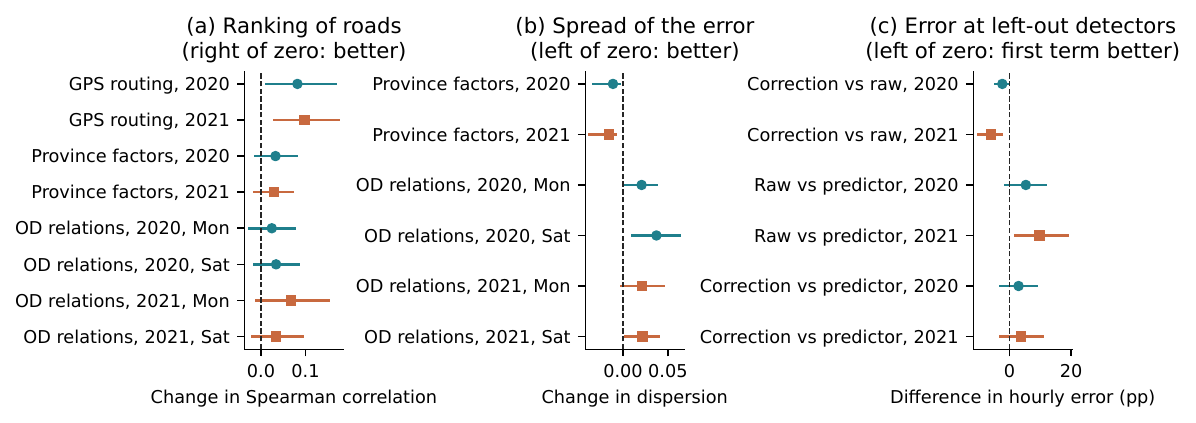}
\caption{\textbf{The marginal information of each source.} Each row is one comparison on data not used to set the values: the marker is the estimated change and the bar its 95\% bootstrap interval. A bar that crosses the dashed zero line means that the data cannot establish a change. Blue circles: 2020; orange squares: 2021. For the OD relations, the regional pattern of the factors is kept in the simulation without relations (Supplementary Section~S9). \textbf{(a)} Change in the rank correlation between simulated and observed volumes when the source is added. GPS routing helps in both campaigns; no interval of the OD relations clears zero. \textbf{(b)} Change in dispersion, the spread of the error between roads. The province factors reduce it, and the OD relations do not. \textbf{(c)} Differences in hourly error at left-out detectors. The correction beats the raw simulation, but not the predictor without simulation.}
\label{fig:sources}
\end{figure*}

The contribution of GPS appears when route choice is changed under the same demand (Table~\ref{tab:volumes}, bottom, and Fig.~\ref{fig:sources}a). At detectors outside the province where route choice was set, GPS-informed routing ranks roads better than shortest-distance routing in both campaigns, while the typical error does not change measurably. This gain comes mainly from the use of each road class, and GPS speeds alone add little (Supplementary Section~S8). Repeating the selection on the final model confirms the chosen form.

\begin{table}[htbp]
\centering
\caption{Daily volumes at held-out detectors, raw simulation, with 95\% bootstrap intervals. A volume ratio of one means that the total is exact, a Spearman correlation of one that roads are ranked exactly, and a median $|\log_{10}|$ ratio of 0.15 that the typical road is within about 40\% of its count. Top: all detectors of each set. Bottom: GPS-informed against shortest-distance routing under the same demand, at test detectors outside Forl\`i--Cesena; intervals are for the paired difference.}
\label{tab:volumes}
\begin{tabular}{@{}p{4.6cm}p{5.6cm}p{5.6cm}@{}}
\toprule
Measure & Test half, 2020 (97 detectors) & 2021 (141 detectors) \\
\midrule
Volume ratio & 1.00 [0.89, 1.11] & 1.10 [1.01, 1.20] \\
Spearman correlation & 0.69 [0.56, 0.79] & 0.66 [0.54, 0.74] \\
Median $|\log_{10}|$ ratio & 0.148 [0.113, 0.176] & 0.142 [0.118, 0.174] \\
Within a factor of two (\%) & 81 [73, 90] & 84 [78, 90] \\
\midrule
 & Test half, outside FC (87) & 2021, outside FC (128) \\
Spearman, shortest distance to GPS & 0.63 to 0.71, $+0.08$ [0.01, 0.17] & 0.54 to 0.64, $+0.10$ [0.03, 0.18] \\
Median $|\log_{10}|$ ratio, shortest distance to GPS & 0.151 to 0.147, $0.00$ [$-0.04$, 0.03] & 0.152 to 0.141, $-0.01$ [$-0.04$, 0.03] \\
\bottomrule
\end{tabular}
\end{table}

All results so far describe the raw simulation. We now add the local correction. To test it fairly, we leave out each detector in turn and ask the correction, which uses the other detectors, to predict its count. The correction removes 2.4 percentage points of the error in 2020 and 6.0 in 2021 at detectors it has not seen (Table~\ref{tab:correction}), and the spread between detectors remains larger than the gain. About three detectors in five improve and about one in four gets worse. The 2020 set includes the calibration detectors and the depth of the correction was selected on 2020, so 2021 is the estimate free of that selection, although it also includes 60 calibration detectors (Supplementary Section~S4). The gain grows with the number of detectors used (Supplementary Section~S8).

Again we compare with the predictor built on the other detectors of the same day (Table~\ref{tab:correction} and Fig.~\ref{fig:sources}c). The corrected reconstruction cannot be told apart from it, and each is better at about half of the detectors, although the point estimates favour the predictor; the raw simulation is worse than the predictor in 2021, with an interval that excludes zero. So the correction improves the raw reconstruction but does not outperform this benchmark at individual detectors. The simulation, in addition, gives traffic on all links of the network, which the predictor does not; how accurate it is away from detectors is tested here only at left-out detector locations.

\begin{table}[htbp]
\centering
\caption{Local correction at left-out detectors, depth 3: each detector is predicted by a correction that does not use it, so a lower corrected error is a real gain at a road the correction has not seen. Hourly error pooled over hours, directions and days. Predictor: road class and mean profile from the other detectors of the same day. Intervals: 95\% bootstrap.}
\label{tab:correction}
\begin{tabular}{@{}lll@{}}
\toprule
Measure & 2020 (152 detectors) & 2021 (141 detectors) \\
\midrule
Hourly error, raw (\%) & 42.2 & 47.4 \\
Hourly error, corrected (\%) & 39.8 & 41.4 \\
Hourly error, predictor (\%) & 36.9 & 37.6 \\
Reduction by the correction (pp) & 2.4 [0.2, 4.8] & 6.0 [2.5, 10.3] \\
Raw minus predictor (pp) & $[-1.3, +11.6]$ & $[+1.7, +18.9]$ \\
Corrected minus predictor (pp) & $[-3.0, +8.9]$ & $[-3.1, +10.8]$ \\
Median detector, raw to corrected (\%) & 35.9 to 33.3 & 39.4 to 34.8 \\
Detectors improved / worsened (\%) & 60 / 26 & 62 / 23 \\
Detectors better than the predictor (\%) & 47 & 50 \\
Spearman of volumes, raw to corrected & 0.72 to 0.72 & 0.66 to 0.69 \\
\bottomrule
\end{tabular}
\end{table}

\begin{figure*}[htbp]
\centering
\includegraphics[width=\textwidth]{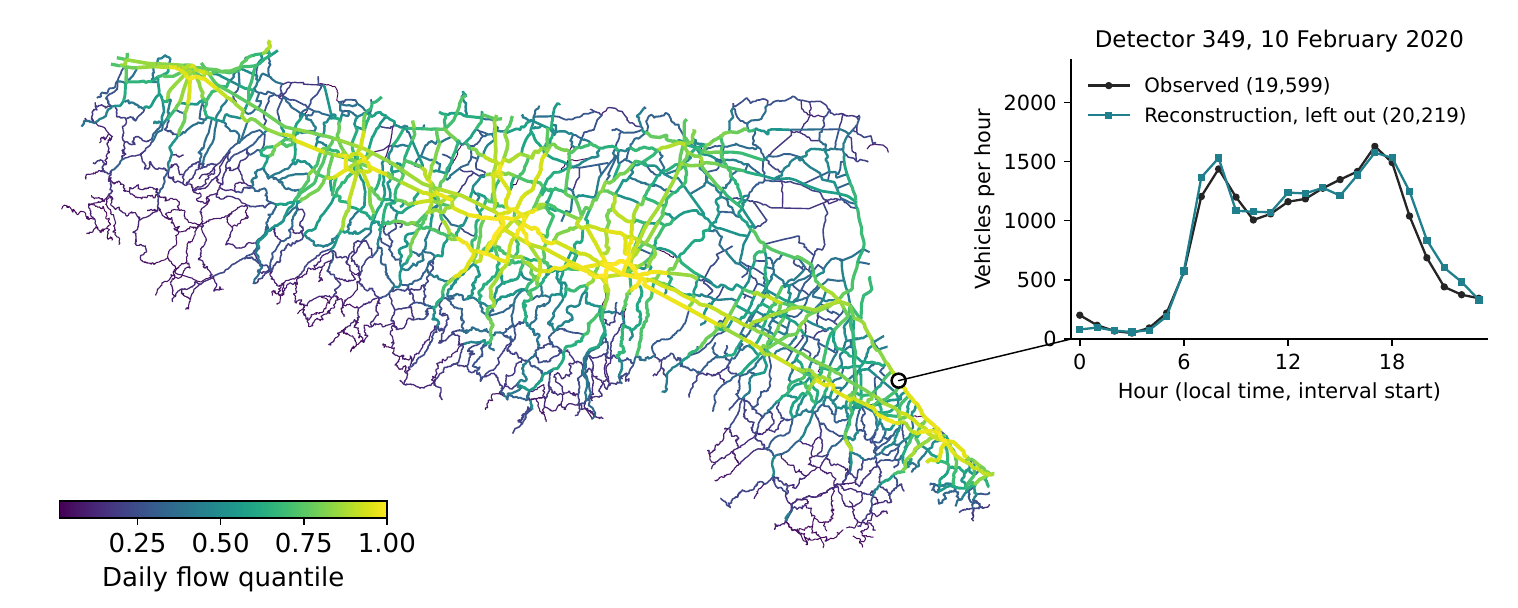}
\caption{\textbf{Reconstructed traffic field.} The simulation and the local correction give traffic on every link, including roads without detectors. Map: reconstructed daily flow on each link, 10 February 2020, with all detectors used. Colour and line width show the quantile of the flow, so yellow marks the busiest roads of the region. Inset: the best case among the 61 detectors complete on that day, detector 349 (circle on the map): hourly counts, both directions, and the reconstruction on its link when that detector is left out of the correction and all others are used; daily totals in parentheses. The error is 9\%, against a median of 35\% over the 61 detectors.}
\label{fig:simulation}
\end{figure*}

With the correction we can draw the full traffic field (Fig.~\ref{fig:simulation}). The motorway axis and the roads around the main cities carry most traffic, and the Apennine roads in the south carry little. The inset shows the best case of the test: with detector 349 left out, the reconstruction on its road follows the daily curve and misses the daily total by about 3\%. This is what the simulator can achieve where it works best (Table~\ref{tab:correction} gives all detectors). Figure~\ref{fig:sources} collects the gains and differences discussed so far, together with the OD relations.

We tested the OD relations in detail in Supplementary Section~S9. Removing them lowers the ranking of roads on the test half, but this also removes the regional pattern of the province factors. With that pattern kept, no gain in ranking can be established, and the spread of errors is larger with the relations (intervals excluding zero in three of four comparisons). Destinations rearranged at random make trips far too long and tend to raise the spread, while destinations rearranged at the same distances cannot be told apart from the real ones. The value of specific origin--destination pairings could not be determined from link-count observations with the statistical power available in this study: in the registered test the power was at most 0.11, far below the 80\% it required. This test is therefore inconclusive: it shows neither that the pairings improve the reconstruction nor that they are unnecessary.

\subsection{Structure of the remaining error}
\label{sec:single_roads}
\begin{figure*}[htbp]
\centering
\includegraphics[width=\textwidth]{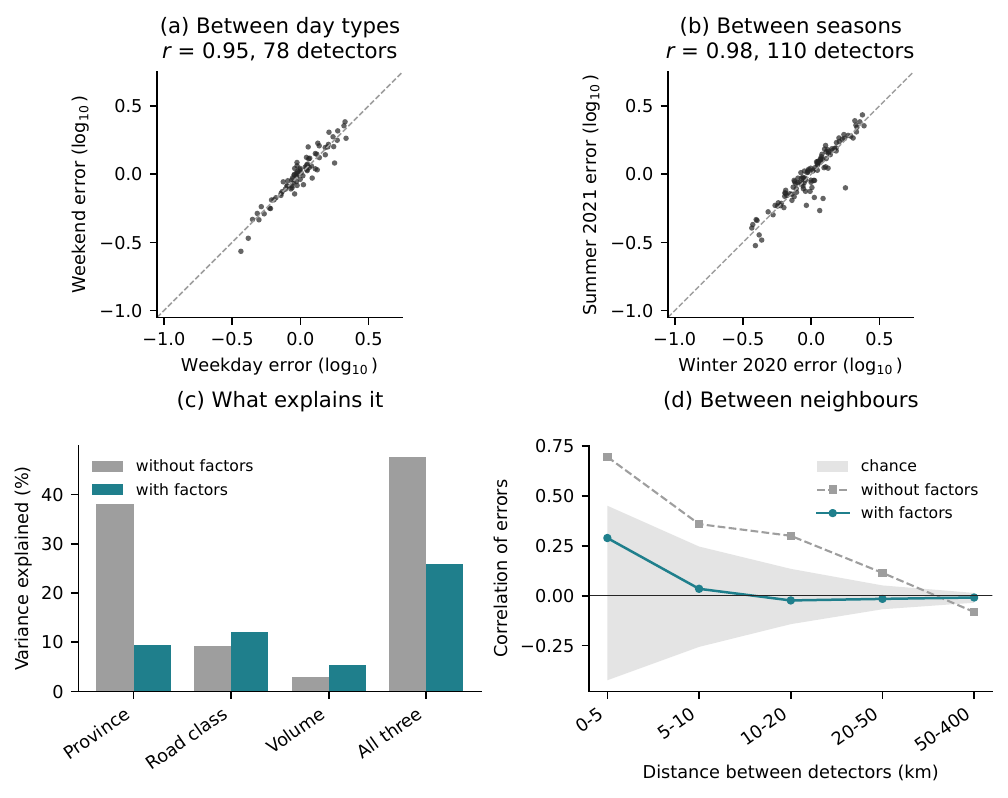}
\caption{\textbf{The remaining error is stable and specific to each road.} The error is $\log_{10}$(simulated/observed) of daily volume, so zero is perfect and 0.3 is about a factor of two. \textbf{(a)} Error of each calibration detector on weekdays against weekends. \textbf{(b)} Error of each detector in winter 2020 against summer 2021. Points close to the dashed diagonal have the same error in both conditions. \textbf{(c)} Share of the variance of the error across calibration detectors explained by province, road class, observed volume and all three together, on the Monday, without and with province factors. \textbf{(d)} Correlation between the errors of pairs of detectors, by distance. The shaded band is the 95\% range expected by chance, from permutations with the factors; a point inside it shows no detectable similarity between neighbours.}
\label{fig:single}
\end{figure*}

With the regional level, the ranking and the timing in place, we can ask what kind of error remains at each road. Figure~\ref{fig:single}a,b shows that it is not noise but a stable property of the road. A road that the simulator overestimates on weekdays is overestimated by about the same amount on weekends ($r=0.95$), and the same holds between winter 2020 and summer 2021 ($r=0.98$). The day-to-day variation of the counts is about ten times smaller than this error. Part of the stability between seasons is expected, since both campaigns use the same demand; it also means that the counts on each road changed little relative to the simulated ones.

Figures~\ref{fig:single}c and \ref{fig:single}d show what the province factors did to this error. Without them, province was its largest known component and nearby detectors had similar errors. With them, that regional component has mostly gone and nearby errors are no more alike than chance would give. Together, province, road class and volume now explain about a quarter of the variance. Misplaced detectors do not explain the rest: comparing their coordinates with the road geometry, we flagged 5 of 292 as doubtful, and dispersion remains large without them (Supplementary Section~S5). Of the other road properties we tested out of sample (nearby population and activities, centrality, parallel roads, density of roads not in the network) and of route weights from the GPS count of each link, none explains the error in a way that carries over between provinces (Supplementary Section~S8). What remains is a stable property of each road, which may reflect how demand reaches the road locally, and which these data cannot separate further. Its stability suggests that local information could address it, as the correction already does where detectors exist.

\subsection{Formal tests and contribution of each source}
\label{sec:tests}
Finally, we test each property on held-out data (Table~\ref{tab:tests}). The overall level is equivalent to the observed one within the chosen margin, the ranking of roads exceeds chance even when road class is held fixed, and the timing beats a flat profile on nearly all links of Forl\`i--Cesena. A regression across detectors gives a slope close to one, with much of the variance left to the single-road error described above. The province interval of the level test is based on few provinces, each of which received its own factor, so it is only partly held out.

\begin{table}[htbp]
\centering
\caption{Final tests on data not used to set the model values, raw simulation. The volume ratio passes if its interval lies within 0.80--1.25; the two Spearman tests ask whether the ranking of roads is better than chance, overall and within road class; the timing test whether the daily timing is better than a flat profile. Intervals for the volume ratio are 90\%; for the slope, 95\%. FC: Forl\`i--Cesena. The smallest margin of the volume ratio that passes in both campaigns is about 1.20. Holm-adjusted $p<0.001$ for the two Spearman tests. In the timing test the simulation beats the flat profile on more than 99\% of links. The 2021 set includes 60 calibration detectors.}
\label{tab:tests}
\begin{tabular}{@{}p{5.8cm}p{4.4cm}p{4.4cm}@{}}
\toprule
Test & Test half, 2020 (97 detectors) & 2021 (141 detectors) \\
\midrule
Volume ratio, resampling detectors & 1.00 [0.91, 1.09]: pass & 1.10 [1.02, 1.18]: pass \\
Volume ratio, resampling provinces & [0.86, 1.13]: pass & [0.97, 1.20]: pass \\
Spearman above chance & 0.69, $p<0.001$ & 0.66, $p<0.001$ \\
Spearman within road class & $p<0.001$ & $p<0.001$ \\
Timing against flat, FC links (error, pp) & 1.08 against 2.72, $p<10^{-70}$ & 1.21 against 2.60, $p<10^{-70}$ \\
Slope, log--log volumes ($R^2$) & 0.94 [0.72, 1.18] (0.19) & 0.86 [0.70, 1.05] (0.17) \\
\bottomrule
\end{tabular}
\end{table}

Figure~\ref{fig:sources} and Supplementary Table~\ref{tab:sources} collect the effect of each source on data not used to set it. Phone data give the timing and, with detector scaling, the level; GPS improves the ranking of roads; detectors also bring a local gain through the correction. The OD tables add the short-range structure of trips; the value of specific pairings beyond it remains undetermined, because the test had too little power.

\section{Discussion}
\label{sec:discussion}
With eleven demand-level parameters (two by day type, nine by province), the simulator reaches the observed overall level and timing, ranks roads well above chance, and the roles of GPS, the detectors and the demand profiles can be measured on held-out data, whereas that of specific OD pairings is not resolved. The province factors are empirical: a factor of 1.74 or 0.51 cannot be read as a difference in phone coverage or car use without further evidence, and the factors would have to be estimated again for another period or region.

The OD tables need careful reading, because our tests answer narrower questions than "do OD data have value?". Destinations rearranged at random make trips far too long and tend to worsen the model, so the tables carry useful information on trip length. Random ends drawn near the known end supply much of the same information, which may be why removing the relations does not raise dispersion; this explanation is a hypothesis. The value of specific origin--destination pairings could not be determined from link-count observations with the statistical power available in this study. Just over a quarter of the trips between municipalities with known ends keep their relation, and link counts respond little to pairings: at the 25-km radius of the model, the change at detectors is below the variation between seeds. This is not evidence that pairings have no value, and they may be useful in applications that need trips rather than link volumes. In the same way, a value of $k=1.5$ cannot be read as the inverse of the operator's coverage.

The simulation ranks the test half better than a predictor without simulation, with a similar typical error at the detectors; at left-out detectors the predictor has the lower point error even after the local correction. Unlike that predictor, the simulation also provides routes and traffic on roads without detectors, but its accuracy there is inferred only from left-out detector locations, and detectors are not a random sample of roads.

Three findings may apply beyond this region, as hypotheses for other settings. Timing needs a benchmark, since a common daily profile already explains most of the hourly pattern. A few scale parameters can reach the overall level, so the error at single roads should be reported along with it. Removing a source can mislead when other parameters act through it, so its effect should be tested with those parameters kept, and with the marginals preserved. The values we report are marginal to this model and to the order in which we built it, and they are not an exchangeable decomposition. They also concern ordinary traffic.

None of the fitted values should be carried to another region or period: the demand scale, the province factors, the class speeds and usage, the radius and the departure shift would all have to be estimated again. What can be carried over is the procedure: a defined role for each source, held-out data for each role, predictors without simulation as benchmarks, and the effect of each source measured with the parameters that act through it held fixed. Whether other regions show the same pattern of contributions is an open question.

Some boundaries of these results should be kept in mind. GPS comparison covers one province, which also informs route choice. Several routes per relation, shorter walks and a smaller radius improved the ranking of roads in exploratory tests but did not reduce dispersion (Supplementary Section~S8), so the reach of the relations remains limited by the structure of the model. The experiments on the OD relations are exploratory and rest on one simulated Monday and Saturday per campaign. The test half and 2021 took part in the decision on the province factors and had been seen when the final configuration was fixed, so the validation results depend in part on model-development decisions. The 2021 evaluation uses demand measured in early 2020 and a scale fixed on winter counts, so it tests how the spatial pattern persists rather than the demand of that season. No detector lies on a motorway, and roads without detectors are assessed only through left-out detector locations. Several settings (trip radius, cost weights, origin spreading) were chosen on the calibration half without a demonstrated gain on held-out detectors. The largest remaining error is the level at individual roads, so aggregate agreement should not be read as accuracy at a given road.

This paper establishes what the simulator does for ordinary traffic: regional level, ranking of roads and timing. The simulator produces hourly traffic estimates on all roads of the network, including those without detectors, while the available observations establish their accuracy only at detector locations. These estimates come from the simulation; the local correction is a statistical adjustment near detectors, which moves no agents and does not keep flows consistent between links beyond its depth, so corrected and simulated values should be kept apart in later use. The simulator can therefore serve as a computational basis for later work on road closures, demand perturbations and network resilience. We make no claim here about those applications: skill on ordinary traffic does not imply skill under a change, and they need their own assessment on observed changes.

\section{Conclusions}
\label{sec:conclusions}
We have built a regional traffic simulator from mobile-phone demand, detector counts and GPS, in which each source has a defined role, and measured what each source adds on data not used to set each step. On held-out detectors the simulator reaches the overall level (volume ratio 1.00 in winter 2020 and 1.10 in summer 2021),
reproduces the observed ranking of road volumes better than a road-class baseline, and follows the daily timing, which comes mostly from the demand profiles themselves. Detector scaling sets the level, GPS-based route weights improve the ranking, and the local correction reduces the error at unseen detectors, although it does not outperform a predictor without simulation at individual detectors. These held-out sets also informed some modelling choices, and the 2021 set shares 60 detectors and the 2020 demand with the calibration, so the evaluation is held out rather than fully independent. What remains uncertain is the error at single roads, which is stable, specific to each road and comparable to that of simple predictors; the accuracy on roads without detectors, motorways in particular; the value of specific origin--destination pairings, whose test was inconclusive for lack of power; and the use of the simulator for other periods, regions or scenarios, which needs its own assessment on observed changes.

\section*{Supplementary material}
The Supplementary Material describes the data layers and the demand classes (S1), the demand scale and the changes tested on the calibration detectors (S2), the simulation model and the route-choice variants (S3), the detector correction and its sensitivity to depth (S4), additional results (S5), the final configuration and the reproduction of all results (S6), further details of data and methods (S7), the experiments on the final model (S8), and the value of the OD relations together with an exploratory comparison with detector counts of winter 2023, which is not an independent replication (S9).

\section*{Abbreviations}
GPS, Global Positioning System; MAE, mean absolute error; MTS, regional traffic monitoring system of Emilia-Romagna; OD, origin--destination; pp, percentage points.

\section*{CRediT authorship contribution statement}
\textbf{Federico Bellisardi:} Conceptualization, Methodology, Software, Formal analysis, Investigation, Visualization, Writing -- original draft.
\textbf{Chiara Mizzi:} Software, Data curation.
\textbf{Alessandro Fabbri:} Software, Data curation.
\textbf{Davide Micheli:}  Resources, Data curation.
\textbf{Aldo Vannelli:} Resources, Data curation.
\textbf{Armando Bazzani:} Conceptualization, Methodology, Supervision.

\section*{Declaration of competing interest}
C.M. and A.F. are employed by Mindicity, which contributed to the development of the modelling framework during their period at the University of Bologna. D.M. and A.V. are affiliated with TIM -- Telecom Italia S.p.A., which provided the mobile-phone data. The other authors declare no competing interests.

\section*{Funding}
This study was carried out within the MOST -- Sustainable Mobility National Research Center and received funding from the European Union NextGenerationEU under the Italian National Recovery and Resilience Plan (PNRR, Mission 4, Component 2, Investment 1.4; D.D. 1033 of 17/06/2022, CN00000023). F.B. received support from the Italian National Operational Programme on Research and Innovation 2014--2020 (CCI 2014IT16M2OP005), FSE REACT-EU, Actions IV.4 and IV.5. The funders had no role in the study design, in the collection, analysis and interpretation of data, in the writing of the article or in the decision to submit it for publication.

\section*{Data availability}
The mobile-phone OD tables (TIM) and the Viasat GPS data are proprietary and were used under agreements that do not allow redistribution; requests for access should be addressed to the data providers. MTS loop-detector counts are published by Regione Emilia-Romagna under a CC BY licence \cite{mtsdata}. Road classes are derived from OpenStreetMap (\copyright{} OpenStreetMap contributors, ODbL). The road graph and the aggregate results behind the figures and tables are included in the code repository; link-level metrics derived from the GPS data are not distributed. The simulation configurations derive from the mobile-phone OD tables and are not distributed. The simulator \texttt{minimocas}, the detector correction and all analysis scripts are available in the code repository \cite{repository}. Supplementary Section~S6 lists the final configuration and the steps that reproduce the results.

\section*{Acknowledgements}
We thank TIM -- Telecom Italia S.p.A.\ for providing access to the MDT dataset, Viasat for providing the GPS data, the Mobility and Viability Section of the Emilia-Romagna Region for providing the MTS traffic detector data, and Stefano Grandi, Claudio
Domenichini, and Elena Grossi for useful discussions.

\bibliographystyle{elsarticle-num}
\bibliography{references}

\clearpage
\newgeometry{margin=1.8cm}
\setcounter{section}{0}
\setcounter{subsection}{0}
\setcounter{table}{0}
\setcounter{figure}{0}
\setcounter{equation}{0}
\renewcommand{\thesection}{S\arabic{section}}
\renewcommand{\thesubsection}{S\arabic{section}.\arabic{subsection}}
\renewcommand{\thetable}{S\arabic{table}}
\renewcommand{\thefigure}{S\arabic{figure}}
\renewcommand{\theequation}{S\arabic{equation}}
\renewcommand{\theHsection}{S\arabic{section}}
\renewcommand{\theHsubsection}{S\arabic{section}.\arabic{subsection}}
\renewcommand{\theHtable}{S\arabic{table}}
\renewcommand{\theHfigure}{S\arabic{figure}}
\renewcommand{\theHequation}{S\arabic{equation}}
\setlength{\tabcolsep}{4pt}
\renewcommand{\arraystretch}{1.12}
\begin{center}
{\Large Supplementary Material}
\end{center}
\noindent This document gives further detail on the data layers, the simulation model, the route-choice variants and the detector correction described in the main manuscript. It also reports additional results and lists the commands that reproduce them.

\section{Data layers}
\label{sec:data-layers}

\begin{table}[htbp]\centering
\caption{Data layers and their role in the reconstruction.}
\label{tab:layers}
\begin{tabularx}{\textwidth}{@{}l>{\raggedright\arraybackslash}X>{\raggedright\arraybackslash}Xl@{}}\toprule
Layer & Content & Role & Period \\\midrule
Mobile-phone OD (TIM) & hourly device movements between census areas, aggregated to municipalities and provinces & demand & 13 January--23 February 2020 \\
Road graph & 3,925 directed links, 2,548 nodes, turn restrictions, derived from OpenStreetMap & network & -- \\
OSM road classes & motorway to unclassified, nearest ways within 30~m & route-choice parameters & extract of October 2026 \\
MTS loop detectors & 15-min counts per station and direction & demand scale and demand settings (half of the detectors of February 2020); local correction; volume comparison & Feb 2020, Jun--Jul 2021 \\
Viasat GPS, link counts & hourly probes per matched link, Forl\`i--Cesena & profile comparison & 10--17/02/2020, 28/06--04/07/2021 \\
Viasat GPS, raw traces & private-vehicle fixes in Forl\`i--Cesena & speeds and class usage for route choice & same periods \\
\bottomrule\end{tabularx}
\end{table}

Demand is generated in five classes, after removing the cells with the code 000000000. Agents have an observed source and destination between the 50 most active municipalities (23\% of the agents on a Monday) or between them and the provinces of neighbouring regions (5\%), an observed source and a random destination (20\%), a random source and an observed destination (20\%), or a random source and destination (31\%). This last class represents the activity not attributed to the selected relations, including trips within municipalities outside the 50. Agents with a random end therefore account for 72\% of the 1.99 million agents generated on a Monday, after the province factors. The hour convention of the OD tables is discussed in Section~\ref{sec:further}.

We obtained the MTS counts of 2021 from the monthly open-data files of Regione Emilia-Romagna (dataset DS-1944, CC BY), summing all vehicle classes. The time labels of these files are in local time; converted to UTC, they reproduce the counts of the February 2020 archive used for 2020 (ratio of totals 0.995 on 12 February 2020).

\begin{figure}[htbp]
\centering
\includegraphics[width=\textwidth]{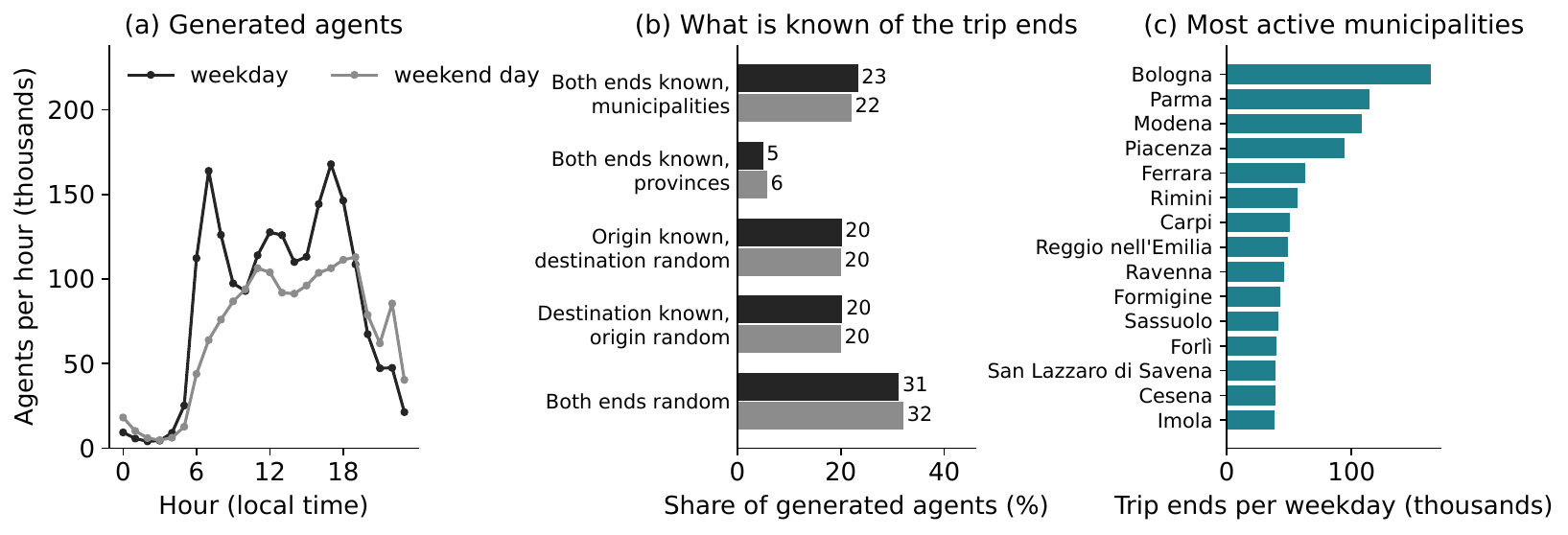}
\caption{\textbf{Simulated demand.} \textbf{(a)} Agents generated per hour for a weekday and a weekend day, after scaling by $k$ and the province factors: weekdays show morning and evening peaks, and weekends a single broad rise through the day. \textbf{(b)} Shares of agents by what is known of their trip ends; "provinces" are relations with provinces of neighbouring regions. More than a quarter of agents have both ends known, and the largest group has both ends drawn at random. \textbf{(c)} Municipalities with the most trip ends on a weekday.}
\label{fig:demand}
\end{figure}

Figure~\ref{fig:demand} shows the simulated demand by hour and by what is known of the trip ends.

\section{Demand scale}
\label{sec:scale}

The configuration generator inherited from the operational platform rescaled individual source--destination pairs with ratios between observed and simulated counts at detectors. Because these ratios used the counts of the simulated day itself, the simulation was not independent of the detectors. We switched this rescaling off and instead multiply the demand by a factor $k$ for each day type (main text, Methods). Earlier scans and a fit by demand class, made on a previous version of the engine and demand, are not reported.

The same generator computed the relations with neighbouring provinces from all OD cells, while the classes with random ends were computed from the same cells without removing them, so those trips were generated twice: 63,532 agents on a Monday (4.4\%) and 68,388 on a Saturday (5.0\%), before scaling. We corrected this and repeated all simulations and selections, with the decision to report the corrected model fixed before the runs. All results in this paper use the corrected model. On the previous version, the selected radius and departure shift were the same, $k$ was 1.4 and 1.2, and the volume ratio, Spearman correlation and dispersion on the detectors of 2021 were 1.10, 0.66 and 0.20, as in the corrected model.

\begin{figure}[htbp]
\centering
\includegraphics[width=\textwidth]{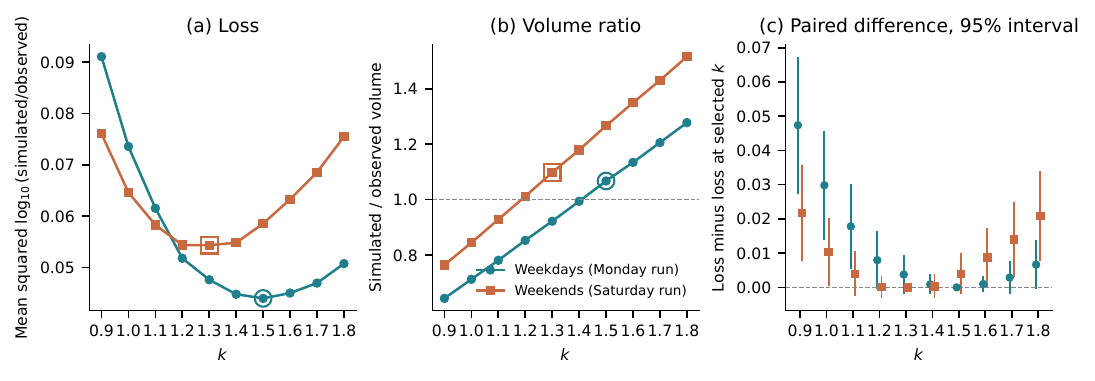}
\caption{\textbf{Choice of the demand scale $k$.} Calibration detectors, 25-km radius, without province factors; 83 detectors on weekdays (Monday run) and 95 on weekends (Saturday run). \textbf{(a)} Mean squared $\log_{10}$(simulated/observed) of daily volumes; large open markers: selected $k$. \textbf{(b)} Ratio of total simulated to total observed volume. \textbf{(c)} Difference in loss from the selected $k$, with 95\% paired bootstrap intervals over detectors (2,000 resamples, seed 20261047).}
\label{fig:k_choice}
\end{figure}

The grid of $k$ was evaluated on the calibration detectors at the 25-km radius, without province factors, which leave $k$ unchanged (Fig.~\ref{fig:k_choice}). The selected values, 1.5 on weekdays and 1.3 on weekends, are the minima of the loss and give volume ratios of 1.07 and 1.10. The minima are flat: the neighbours cannot be told apart on the loss ($k=1.4$ $[-0.002, +0.004]$ and $k=1.6$ $[-0.001, +0.003]$ on weekdays; $k=1.2$ $[-0.003, +0.003]$ and $k=1.4$ $[-0.003, +0.004]$ on weekends). The selected value is the best in 52\% of the bootstrap resamples on weekdays and 20\% on weekends, where $k=1.2$ (44\%) and $k=1.4$ (33\%) are as often or more often the best. The intervals exclude $k\le 1.1$ on weekdays and $k\le 1.0$ and $k\ge 1.6$ on weekends.

Table~\ref{tab:rounds} lists the changes tested on the calibration detectors, each against the previous configuration. Each rule was fixed before its runs and required the 95\% interval to lie below zero. The first two rows and the trips within the 50 municipalities were structural choices tested on an earlier configuration, with the code 000000000 still in the data, and were not reopened; the others come from the final configuration.

\begin{table}[htbp]\centering
\caption{Changes tested on the calibration detectors (the first two rows and the intra-municipal row: earlier configuration). Difference in the standard deviation of $\log_{10}$(simulated/observed), new minus reference; for the departure shift, difference in the pooled hourly error (pp). Intervals: 95\% paired bootstrap over detectors.}
\label{tab:rounds}
\begin{tabular}{@{}p{6.6cm}rlll@{}}\toprule
Change & Difference & Interval & Rule met & Kept \\\midrule
OD destination drawn around the declared one (correction of a defect) & +0.0047 & [$-0.0037$, +0.0123] & no & yes \\
Origins spread around the source point & $-0.0187$ & [$-0.0421$, +0.0059] & no & yes \\
Departures 45 min earlier, Monday (pp) & $-4.05$ & [$-4.99$, $-3.17$] & yes & yes \\
Departures 45 min earlier, Saturday (pp) & +0.01 & [$-0.87$, +0.96] & no & -- \\
Intra-municipal trips added & +0.0105 & [$-0.0011$, +0.0221] & no & no \\
$k$ by day type at 20 km, against common $k=1.5$ & $-0.0029$ & [$-0.0061$, +0.0002] & no & yes \\
Radius 25 km against 20 km & $-0.0006$ & [$-0.0096$, +0.0086] & no & yes \\
\bottomrule\end{tabular}\end{table}

The destination change corrects a verified defect, so we kept it regardless of the result. The origin spreading has no demonstrated effect. The factors by day type bring the level of both day types close to one without measurably changing the dispersion. The radius and $k$ follow the minimum of the loss, so the 25-km radius is the selected value of that rule, without a demonstrated gain over 20~km.

\paragraph{Province factors.} The factors were set on the calibration half in two rounds. Dispersion and Spearman correlation are 0.222 and 0.66 without them, 0.193 and 0.72 after the first round, and 0.185 and 0.74 after the second. The factors are 1.74 (Piacenza), 1.42 (Parma), 0.51 (Reggio Emilia), 1.19 (Modena), 0.90 (Bologna), 1.48 (Ferrara), 0.82 (Ravenna), 0.70 (Forl\`i--Cesena) and 0.55 (Rimini). The evaluation uses the rule stated in the main text, which is met on the test half and in 2021; the dispersion there includes a detector without simulated traffic.

Table~\ref{tab:province_effect} gives the effect of the factors on all three sets of detectors.

\begin{table}[htbp]
\centering
\caption{Effect of the province factors, without and with them. Lower dispersion and median error, higher Spearman correlation and a volume ratio closer to one are better; an interval that excludes zero marks a change the data can establish. Dispersion: standard deviation of $\log_{10}$(simulated/observed) of daily volumes across detectors; the test-half values include detector 153, which has no simulated traffic. Median error: median absolute $\log_{10}$ ratio per detector and day type. Intervals: 95\% paired bootstrap of the change.}
\label{tab:province_effect}
\begin{tabular}{@{}lrrrrrr@{}}
\toprule
 & \multicolumn{2}{c}{Calibration half} & \multicolumn{2}{c}{Test half, 2020} & \multicolumn{2}{c}{2021} \\
Measure & without & with & without & with & without & with \\
\midrule
Dispersion & 0.222 & 0.185 & 0.500 & 0.489 & 0.415 & 0.399 \\
\quad change, interval & \multicolumn{2}{r}{$[-0.052, -0.022]$} & \multicolumn{2}{r}{$[-0.034, -0.003]$} & \multicolumn{2}{r}{$[-0.037, -0.008]$} \\
Spearman correlation & 0.66 & 0.74 & 0.63 & 0.66 & 0.63 & 0.66 \\
\quad change, interval & \multicolumn{2}{r}{$[0.02, 0.13]$} & \multicolumn{2}{r}{$[-0.01, 0.08]$} & \multicolumn{2}{r}{$[-0.01, 0.07]$} \\
Median error & 0.151 & 0.101 & 0.180 & 0.148 & 0.177 & 0.142 \\
Volume ratio & 1.08 & 1.02 & 1.06 & 1.01 & 1.19 & 1.10 \\
\bottomrule
\end{tabular}
\end{table}

\section{Simulation model and route-choice variants}
\label{sec:routing}

Agents move at the speed of their link, $v_i$, which is reduced linearly with density $\rho_i$ below the critical density and set to a critical speed above it. Link speeds are read in km/h from the cartography and assigned by road class from GPS (Table~\ref{tab:classes}). Agents with a random end have a trip-time budget drawn uniformly between 900 and 5,600~s and choose each next link from the transition weights, so a known end sets where they start and does not steer them, while source--destination agents follow the best-path matrix of their type. When the turn-restricted search finds no feasible next node for the cost of their type, these agents use the shortest-distance path matrix. The simulator is deterministic (fixed random seed).

Table~\ref{tab:routing} lists the variants simulated for 10 February 2020. They were run with the demand of the earlier generator, before the demand scale was revised; the selection uses only GPS counts in Forl\`i--Cesena. Speeds alone (second to fourth rows) do not change the distribution of traffic over road classes, because most agents follow transition probabilities rather than best paths. Transition weights by road class (last three rows) raise the share of main roads and the agreement with GPS counts in Forl\`i--Cesena, and the selected variant combines the usage cost of Eq.~\eqref{eq:route_cost} with these weights. Table~\ref{tab:routing_volumes} compares shortest-distance and GPS-informed route choice over the full campaigns with the scaled demand, and Table~\ref{tab:routing_shape} shows that route choice hardly changes the hourly profiles.

\begin{table}[htbp]\centering
\caption{Road-class properties from private-vehicle GPS traces in Forl\`i--Cesena, February 2020. Median off-peak speed of links with at least 50 matched points (interquartile range in brackets) and median daily probes per covered link, both applied to all links of each class in the region.}
\label{tab:classes}
\begin{tabular}{@{}lrrrr@{}}
\toprule
Road class & Links (speed) & Speed (km/h) & Links (usage) & Probes per day \\
\midrule
Motorway & 44 & 123 [120--124] & 61 & 130 \\
Trunk & 75 & 83 [76--95] & 76 & 79 \\
Primary & 151 & 55 [46--64] & 160 & 109 \\
Secondary & 245 & 51 [46--60] & 248 & 36 \\
Tertiary & 193 & 53 [43--59] & 227 & 21 \\
Unclassified & 42 & 49 [36--54] & 57 & 14 \\
\bottomrule
\end{tabular}
\end{table}

The variants of Table~\ref{tab:routing} correspond, in Eq.~\eqref{eq:route_cost}, to shortest distance $(\alpha,\alpha_s)=(1,0)$, travel time $(0,1)$, mixed $(0.5,0.5)$ and usage cost $(0.5,0)$. The final configuration uses $(0.5,0.25)$ for all agents.

The cost weights were set in a further step. We simulated a grid of $(\alpha,\alpha_s)$ with all else fixed and compared it with $(0.5,0)$ on the 100 calibration detectors, for 10 and 15 February 2020 (Table~\ref{tab:routing_grid}). Under the rule fixed before the results were seen, a variant replaces the reference if the 95\% paired bootstrap interval of the change in dispersion lies below zero. $(0.5,0.25)$ met it, in one of seven variants and without correction for multiplicity; on held-out detectors the interval reached zero in all four comparisons. We applied the rule and repeated all later steps with this cost. The earlier selection on GPS is repeated on the final model in Section~\ref{sec:experiments}.

\input{tables/routing_variants}
\input{tables/routing_volumes}
\input{tables/routing_shape}

\begin{table}[htbp]
\centering
\caption{Grid of cost weights on the 100 calibration detectors, 10 and 15 February 2020, without the code 000000000 and before the final choice of cost. Dispersion and Spearman correlation of daily volumes over detector-days; 95\% paired bootstrap interval over detectors of the change in dispersion from the reference.}
\label{tab:routing_grid}
\begin{tabular}{@{}rrrrc@{}}
\toprule
$\alpha$ & $\alpha_s$ & Dispersion & Spearman & Change in dispersion \\
\midrule
0.50 & 0.00 & 0.189 & 0.736 & reference \\
1.00 & 0.00 & 0.187 & 0.738 & $[-0.011, +0.008]$ \\
0.75 & 0.25 & 0.187 & 0.739 & $[-0.009, +0.007]$ \\
0.50 & 0.50 & 0.186 & 0.741 & $[-0.010, +0.005]$ \\
0.25 & 0.75 & 0.184 & 0.739 & $[-0.013, +0.003]$ \\
0.00 & 1.00 & 0.183 & 0.741 & $[-0.017, +0.007]$ \\
0.25 & 0.25 & 0.187 & 0.736 & $[-0.007, +0.002]$ \\
0.50 & 0.25 & 0.184 & 0.746 & $[-0.008, -0.001]$ \\
\bottomrule
\end{tabular}
\end{table}

\section{Detector correction}
\label{sec:correction}

The correction starts from the increment of Eq.~\eqref{eq:coil_residual} on each detector link and hour. The simulated count on a link is the number of distinct agents present on it during the hour, while the detector counts passages at a section; the two coincide for agents that traverse the whole link within the hour. Barrier counts do not enter the correction. Two pairs of detectors (404 and 680; 503 and 685) lie on the same link, and their observations are averaged before the increment is computed. Downstream, the increment of a link is passed to each admissible outgoing link, multiplied by that link's share of the simulated outgoing counts at the node; upstream, the same rule applies to incoming links. Each branch carries its own increment and its own path length. The spreading skips other detector links, which receive their own increment, and stops when the increment is below one vehicle, after three links along the path, at a forbidden turn or at a closed link. Every directed link of the raw simulation is kept in the output, and the hours of the observed table are aligned with the local-time hours of the simulation. In the leave-one-detector-out and depth experiments, all detectors on the left-out link are removed together, so that a co-located detector cannot reproduce the left-out count.

Table~\ref{tab:loo} summarizes the leave-one-detector-out test and Table~\ref{tab:depth} the dependence on the depth of the spreading.

\input{tables/loo}
\input{tables/depth}

\section{Additional results}
\label{sec:additional}

Table~\ref{tab:probes} reports the profile error by number of probes per link. The error decreases with the number of probes for both the reconstruction and the mean detector profile, as expected if part of the difference is sampling noise in the probe counts. Matched links with zero simulated or GPS total are excluded, and link-hour records absent from the GPS extract are counted as zero.

\input{tables/probe_strata}

\begin{figure}[htbp]
\centering
\includegraphics[width=\textwidth]{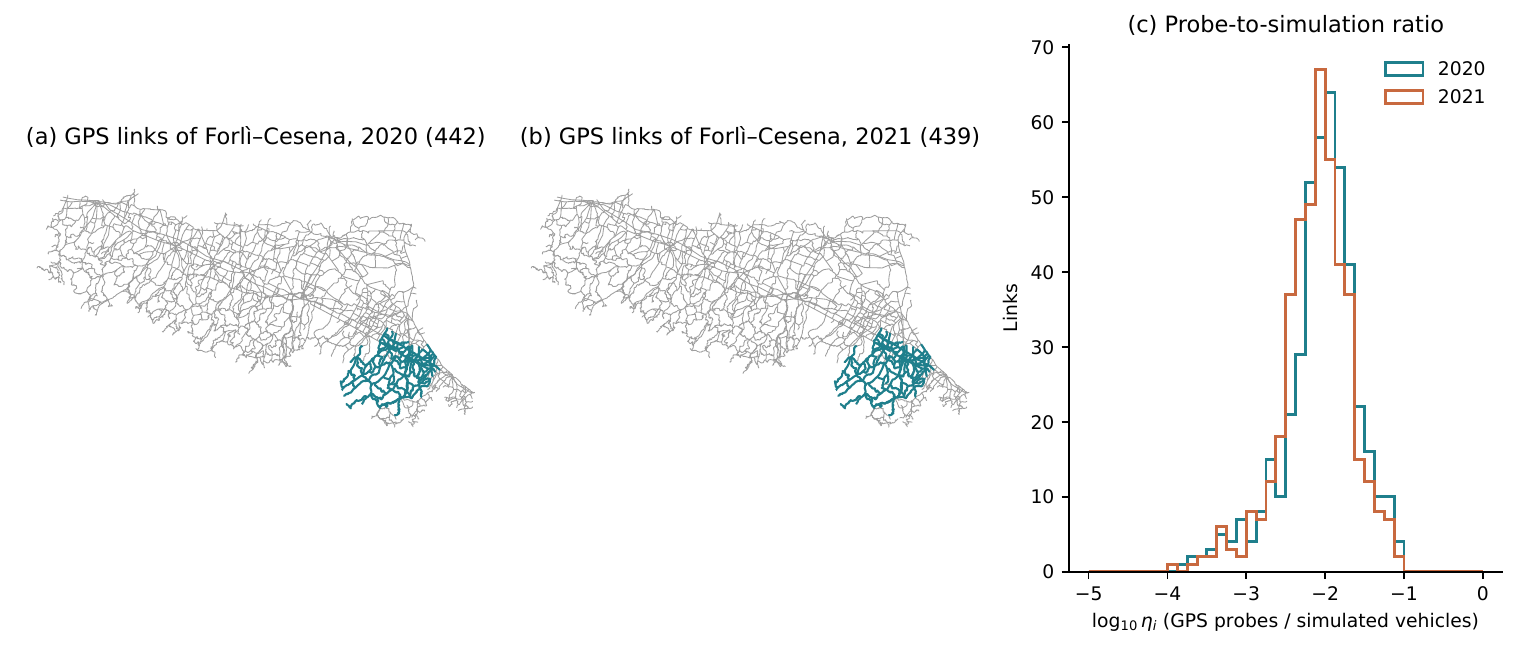}
\caption{\textbf{GPS coverage.} \textbf{(a,b)} Usable GPS-matched links of Forl\`i--Cesena (number in parentheses) over the regional network, 10--17 February 2020 and 28 June--4 July 2021. \textbf{(c)} Distribution of the link ratio $\eta_i$ between probe counts and simulated vehicles, Eq.~\eqref{eq:penetration}. Road network \copyright{} OpenStreetMap contributors.}
\label{fig:viasat}
\end{figure}

\begin{figure}[htbp]
\centering
\includegraphics[width=\textwidth]{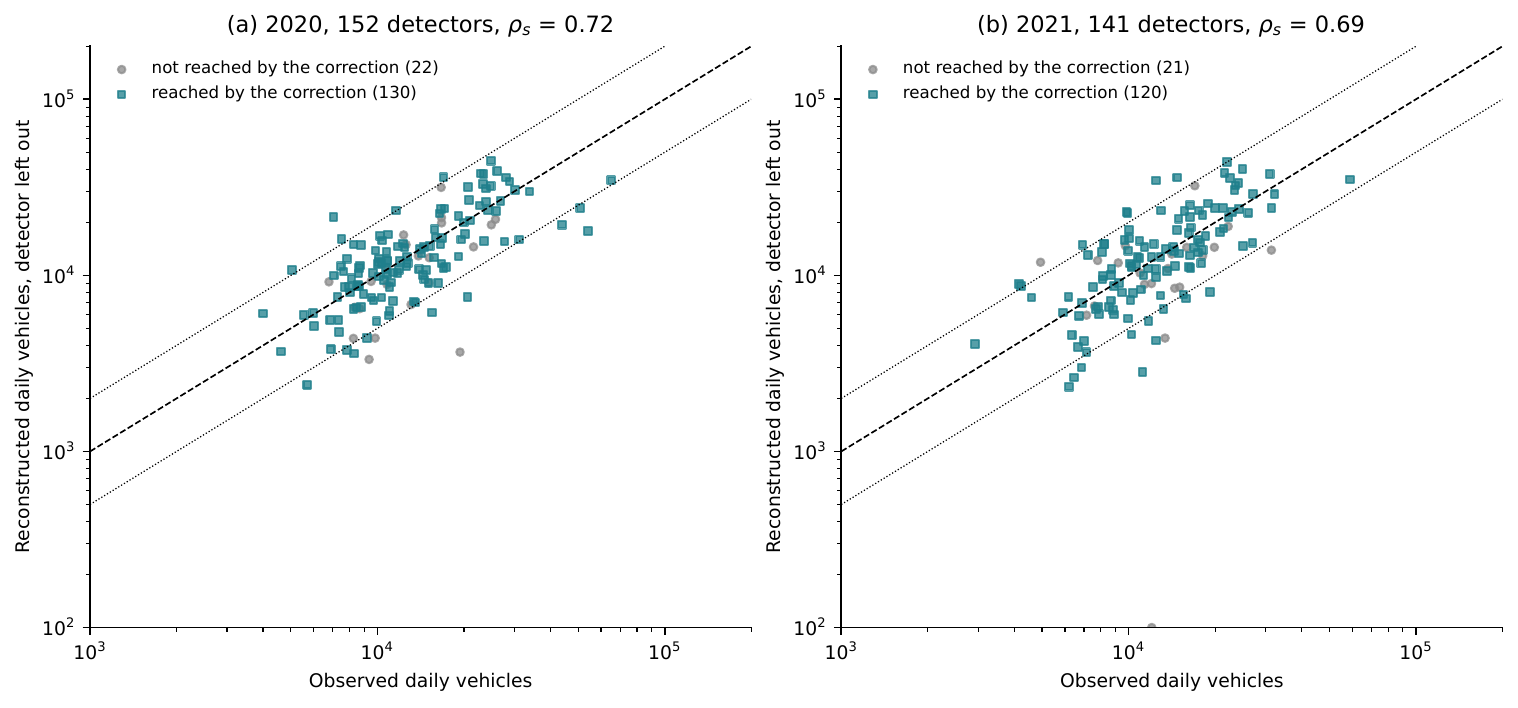}
\caption{\textbf{Daily volumes at left-out detectors.} Observed daily volume of each detector (mean over complete days, both directions) against the reconstruction on its link when the detector is left-out and all others are used, \textbf{(a)} 10--17 February 2020 and \textbf{(b)} 28 June--4 July 2021. Detectors reached by the spreading from other detectors are distinguished from those that are not; $\rho_s$, Spearman correlation. Dashed line: equality; dotted lines: factor of two.}
\label{fig:loo_volumes}
\end{figure}

The ratio between probes $\psi_i$ and simulated vehicles $\phi_i$ on link $i$,
\begin{equation}
\eta_i=\frac{\sum_{d,h}\psi_i(d,h)}{\sum_{d,h}\phi_i(d,h)},
\label{eq:penetration}
\end{equation}
describes how GPS samples traffic (Fig.~\ref{fig:viasat}). Figure~\ref{fig:loo_volumes} shows the daily volumes at left-out detectors.

\begin{figure}[htbp]
\centering
\includegraphics[width=\textwidth]{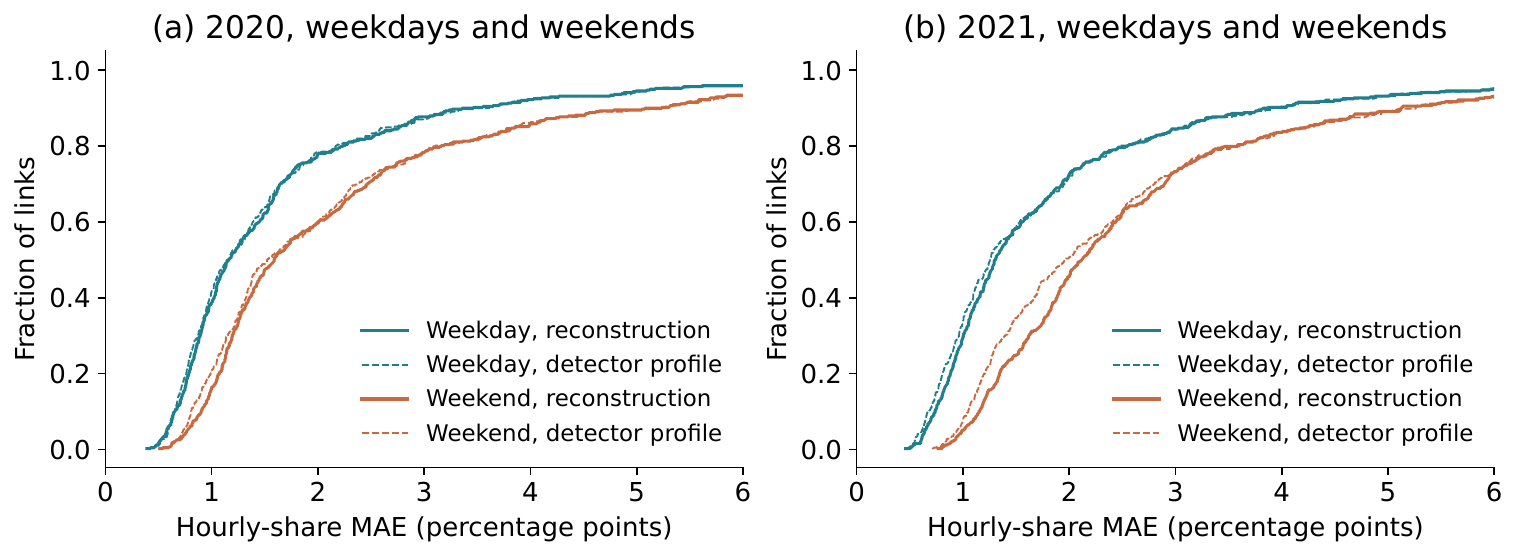}
\caption{\textbf{Timing on weekdays and weekends.} Error of the hourly shares on the GPS-matched links of Forl\`i--Cesena, as in Fig.~\ref{fig:timing_cdf}, for weekdays and weekends separately: reconstruction and mean detector profile. Each curve shows the fraction of links whose error is below the value on the horizontal axis. Weekend curves lie to the right, so errors are larger on weekends. Curves continue beyond the 6~pp limit of the axis.}
\label{fig:daytypes}
\end{figure}

\begin{figure}[htbp]
\centering
\includegraphics[width=\textwidth]{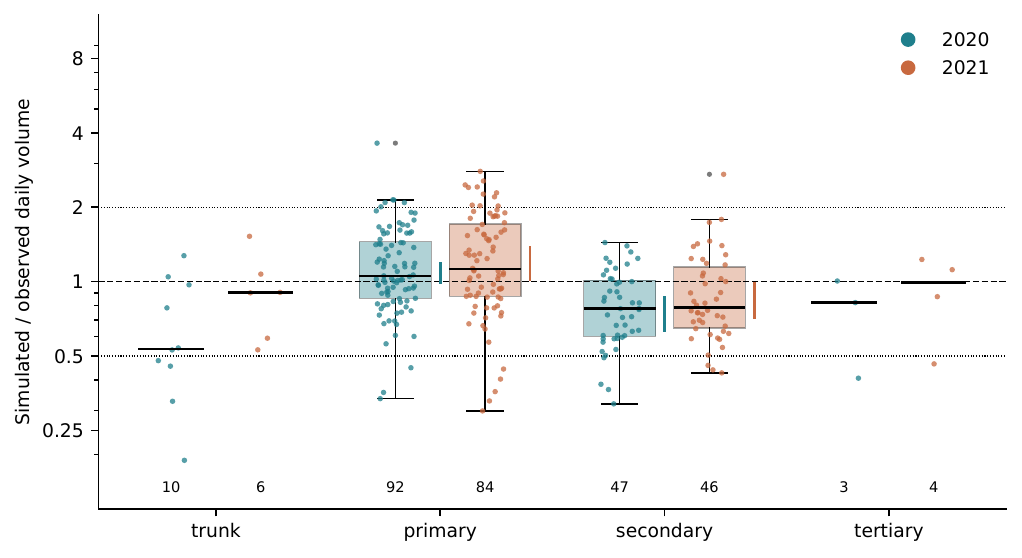}
\caption{\textbf{Simulated volume by road class.} Ratio of simulated to observed daily volume, raw simulation, on the detectors with complete days of the leave-one-out test. Each point is a detector, and the number of detectors is below each group; 106 detectors appear in both campaigns. Black bars: medians. Primary and secondary roads also show quartiles and whiskers at 1.5 interquartile ranges, and the coloured bars beside them are 95\% bootstrap intervals of the median. The dashed line marks perfect agreement and the dotted lines a factor of two. Trunk and tertiary roads have few detectors, and motorways have none. One detector with zero simulated traffic in 2021 is not shown.}
\label{fig:distributions}
\end{figure}

Table~\ref{tab:gps_split} gives the median profile errors behind Fig.~\ref{fig:timing_cdf}, Fig.~\ref{fig:distributions} the simulated volume by road class, and Table~\ref{tab:sources} the effect of each source in numbers, as shown in Fig.~\ref{fig:sources}.

\begin{table}[htbp]
\centering
\caption{How well each estimate reproduces the daily rhythm: median hourly-share error (pp) on the GPS links of Forl\`i--Cesena, all days of each campaign; lower is better. Raw: simulation. Corrected: simulation after the local correction. Mean detector, OD and flat: profiles without simulation, as benchmarks.}
\label{tab:gps_split}
\begin{tabular}{@{}lrrrrrr@{}}
\toprule
Campaign & Links & Raw & Corrected & Mean detector & OD & Flat \\
\midrule
2020 & 442 & 1.08 & 1.05 & 1.03 & 1.14 & 2.72 \\
2021 & 439 & 1.21 & 1.21 & 1.15 & 1.25 & 2.60 \\
\bottomrule
\end{tabular}
\end{table}

\begin{table}[htbp]
\centering
\caption{What each source adds, on data not used to set it. Each row gives a source, its job in the model and the change measured when it is added. Differences are in the sense of adding the source; negative errors are improvements. Intervals are 95\% bootstrap.}
\label{tab:sources}
\begin{tabularx}{\textwidth}{@{}p{4.2cm}p{3.6cm}X@{}}
\toprule
Source & Role & Effect \\
\midrule
Mobile-phone OD, total and timing & agents per interval & hourly-share error on FC links 1.08 and 1.21~pp against 2.72 and 2.60 flat; 0.06 and 0.06~pp lower than the OD profile (median paired difference); mean detector profile 1.03 and 1.15~pp \\
Mobile-phone OD, relations as used (28\% of the trips between municipalities with known ends) & known trip ends & with the factors kept on both sides: Spearman $+0.02$ to $+0.07$, intervals include zero; dispersion $+0.021$ to $+0.037$ (worse), three of four intervals exclude zero. Against random pairs: dispersion $-0.017$ to $-0.039$, all four intervals exclude zero, two of them narrowly. Against pairs with the same distances: no interval excludes zero at 50 municipalities, one of four at 347; value of the pairings not determined (power at most 0.11) \\
Detectors, half & scale $k$ by day type & test ratio 1.00; 1.10 in 2021 \\
Detectors, half (nine values) & province factors & median $|\log_{10}|$ ratio per detector and day type 0.180 to 0.148 on the test half, 0.177 to 0.142 in 2021; dispersion $-0.011$ and $-0.015$ (intervals exclude zero) \\
GPS traces & route weights & Spearman $+0.08$ [0.01, 0.17] and $+0.10$ [0.03, 0.18] outside the province; median error not measurably different; profiles unchanged \\
Detectors, all & local correction & hourly error $-2.4$~pp [$-4.8$, $-0.2$] and $-6.0$~pp [$-10.3$, $-2.5$] at left-out detectors; profile error $-0.01$ and $0.00$~pp on GPS links \\
\bottomrule
\end{tabularx}
\end{table}

\input{tables/od_decomposition}

\paragraph{Further numbers behind the Results of the main text.} \emph{Timing on GPS links of Forl\`i--Cesena.} All days, 2020 and 2021: raw minus mean detector profile $+0.08$ [0.06, 0.09] and $+0.06$ [0.04, 0.08]~pp; corrected minus mean detector profile $+0.01$ [0.00, 0.03] and $+0.02$ [0.01, 0.04]~pp; raw minus OD profile $-0.06$ [$-0.07$, $-0.05$] and $-0.06$ [$-0.08$, $-0.04$]~pp, with the simulation better on weekdays ($-0.10$ and $-0.10$~pp) and worse on weekends ($+0.04$ and $+0.04$~pp). The correction lowers the error by a median of 0.01~pp in 2020 and does not change it in 2021. Because links are spatially correlated, these intervals are optimistic. \emph{Correction.} The reduction at left-out detectors is 2.1 [$-0.6$, 4.8] and 3.1 [1.5, 4.9]~pp on weekdays and weekends in 2020 and 5.9 [1.7, 11.2] and 6.2 [3.7, 9.2]~pp in 2021, on different sets of detectors. Half of the detectors have a corrected error between 23\% and 49\% in 2020 and between 26\% and 55\% in 2021. Cross-validation over 2020 gives 0.9, 1.2, 2.3, 2.5, 3.0 and 3.2~pp at depths 1 to 6; the best depth is the edge of the grid. Detector 349 recorded 19,599 vehicles on 10 February 2020 and the reconstruction with that detector left out gives 20,219 (hourly correlation 0.99). \emph{OD relations.} Removing the relations from the final model, which also removes the regional pattern of the factors, gives Spearman differences of 0.08 [0.02, 0.15] and 0.06 [0.00, 0.12] on the Monday and Saturday of the test half and 0.10 [0.01, 0.20] and 0.05 [$-0.00$, 0.11] in 2021, and dispersion differences of $+0.008$ [$-0.012$, 0.027], $+0.029$ [$-0.000$, 0.059], $+0.009$ [$-0.013$, 0.031] and $+0.010$ [$-0.007$, 0.028]. Spearman correlations with GPS link counts of Forl\`i--Cesena are 0.69--0.73 with and without relations, with all intervals including zero. \emph{Single roads.} The day-to-day deviation of a detector's count from its own mean has a median of 0.017 in $\log_{10}$ units, against a dispersion of 0.167 on the Monday and 0.201 on the Saturday of the calibration half. Without the province factors, province explains 38\% of the variance of the error on the Monday, road class 9\% and observed volume 3\% (48\% together), and detectors less than 5~km apart have correlated errors (0.69, 20 pairs). With the factors, these shares are 9\%, 12\% and 6\% (26\% together), and the correlation is 0.29 below 5~km and $-0.02$ at 10--20~km. The doubtful detectors are two on closed links, one 870~m from its link and two on a road with a different reference; the median distance from the link is 11~m, and on the 193 evaluated detectors not flagged, dispersion remains 0.255 (without province factors, all days).

\section{Final configuration and reproducibility}
\label{sec:reproducibility}

\begin{table}[htbp]\centering
\caption{Final configuration of the model. Every value is fixed before the final simulations.}
\label{tab:final_config}
\begin{tabularx}{\textwidth}{@{}lX@{}}\toprule
Setting & Value \\\midrule
Demand & mean OD of 13 January--23 February 2020 by day of the week, without the code 000000000 and without trips within the 50 municipalities; 50 municipalities; relations with neighbouring provinces counted once \\
Scale $k$ & 1.5 on weekdays, 1.3 on Saturdays and Sundays \\
Province factors & as in Section~\ref{sec:scale}, two rounds \\
Random ends and departures & radius 25~km; origins spread around the source; departures 45~min earlier than the OD label \\
Route choice & $\alpha=0.5$, $\alpha_s=0.25$, GPS class speeds, usage and transition weights by class \\
Network & 3,925 links, 122 closed, unchanged \\
Local correction & depth 3, without travel time \\
Detector split & seed 20261028; 100 calibration, 97 test \\
\bottomrule\end{tabularx}
\end{table}

The code repository contains the simulator and all analysis scripts, and it explains how to build the simulator and set up the Python environment. The script \texttt{hpc/run\_all.sh} runs the simulations and analyses of the final configuration in order; the protocols written before the experiments are in \texttt{model/results/experiments}. The results are reproduced in the following order:
\begin{enumerate}
\item check the hour convention of the OD tables;
\item build the road geometry, the road classes and the GPS link speeds;
\item simulate and select the route-choice variants;
\item build the 2021 detector table from the open-data files;
\item build the demand, run the tests of Table~\ref{tab:rounds}, choose $k$ and the province factors, and run the simulations of the fifteen validation days;
\item run the correction, the GPS comparison and the leave-one-out test;
\item run the volume tests, the comparison of route choices, the depth selection, the predictors without simulation, the tests of the OD relations and the final tests;
\item produce the figures and tables.
\end{enumerate}
Steps 1--7 require the proprietary inputs (mobile-phone OD tables, Viasat link counts and traces), which are not distributed. A simulated day takes 10--30 minutes on an 8-core workstation, depending on the demand factor and on the number of days run in parallel. In a single measured run of the final Monday on a 4-core desktop computer (AMD Ryzen 3 3200G), the simulation took 7.1 minutes of wall-clock time and 2.5~GB of memory and reproduced the link flows of the final run byte for byte.

\section{Further details of data and methods}
\label{sec:further}

\paragraph{Road classes.} Each link was assigned the OpenStreetMap class of the ways nearest to three points along it (within 30~m; Geofabrik extract of October 2026). This gives 709~km of motorways, 326~km of trunk roads, 2,629~km of primary, 4,673~km of secondary and 2,863~km of tertiary roads.

\paragraph{Clock of the OD tables.} The hourly OD labels refer to local time and to the start of the hour. On the days of 10--14 February 2020, the weekday OD profile correlates best with the mean profile of the 90 detectors complete on those days at zero lag (0.96, against 0.94 with the OD profile shifted one hour earlier and 0.74 one hour later), although the mean absolute difference does not separate zero lag from one hour earlier (0.60 and 0.60~pp).

\paragraph{Detector data.} Time stamps are in UTC and are converted to local time; the 2021 open-data release, in local time, reproduces the archive for February 2020. A detector-day is used when both directions are complete for all 96 intervals. The archive contains no zero counts, and more than 95\% of the isolated missing intervals fall before 06:00, which suggests that empty intervals are recorded as missing. The filter therefore favours busy roads (median daily volume of about 17,000 vehicles, against about 8,000 for detector-days with isolated gaps in 2020) and retains fewer detectors on weekdays, whose nights are quieter. Reading isolated gaps (at most 20 intervals, none longer than 4) as zero extends 2020 from 152 to 227 detectors and hardly changes the raw error (39.6\% to 41.1\%, both directions together) or the volume ratio (0.98 to 1.03). 2021 behaves likewise (44.5\% to 44.5\%, ratio 1.11 to 1.11). We did not repeat the leave-one-out test on the extended set.

\paragraph{GPS matching.} GPS fixes with engine on and speed of at least 5~km/h were matched to the nearest link within 20~m whose bearing agrees with the heading within 30$^\circ$. Off-peak speeds are for weekdays, 10:00--16:00. The fallback to the shortest-distance path described in Section~\ref{sec:routing} is rare: 51 next-node decisions in one of the eight simulations, none in the others.

\paragraph{Splits, profiles and bootstrap.} The detector split uses seed 20261028. In the comparison of hourly profiles we exclude links with zero simulated or GPS total and keep only links whose geometry is identical in the graph used for map matching; the mean detector profile is the mean normalized profile of the detector directions complete on all days of the subset. Comparisons between route choices use 2,000 paired bootstrap resamples of detectors (seed 20261029; 5,000 in Table~\ref{tab:routing_volumes}, on all complete detectors). Intervals of the predictors without simulation use 2,000 resamples (seed 20261041). The depth of the correction is selected with seed 20261033, and the effect of the correction on GPS links is summarized by the minimum number of links between a link and any detector link.

\paragraph{More OD relations.} On an earlier configuration, without province factors and with the code 000000000 in the data, $N=100$ raised the Spearman correlation on the calibration detectors from 0.669 to 0.726 (difference interval $[0.017, 0.104]$), but dispersion went from 0.215 to 0.232 ($[-0.003, 0.035]$), so it did not beat the model. The result is hard to interpret, because the share of agents with a random end falls as $N$ grows, which does not separate the information in the relations from the amount of random traffic. Section~\ref{sec:experiments} repeats the test on the final model.

\paragraph{Value of the OD relations, further results.} Table~\ref{tab:od_decomposition} separates the relations from the province factors using runs without the regional pattern; its last block also removes that pattern, and Section~\ref{sec:experiments} gives the complete design. Without factors, the Spearman differences are 0.03--0.07 with intervals that include zero, and dispersion is larger with relations, with all intervals above zero.

\paragraph{Direction of travel.} For each detector we computed the correlation over the 24 hours between the observed and the simulated difference of the two directions. The test is exploratory and changes no setting. Its rule required, on the test half, that the 95\% interval of the difference in median correlation between $N=100$ and the model lie above zero; it is not met on the final model (Section~\ref{sec:experiments}). An error in the mapping between detector and link directions, which we could not check independently, would produce the same weak signal. Comparisons of volumes in this paper therefore sum the two directions.

\section{Experiments on the final model}
\label{sec:experiments}

These experiments followed protocols written before their runs, in the code repository. They change no setting of the model and are exploratory, because the data had been seen. A run with the same engine and configuration reproduced the final Monday simulation byte for byte.

\paragraph{How much of the OD table is kept.} On a Monday the table has 3.15 million trips per day. 17.8\% have the code 000000000 at one or both ends and are removed from the model; the meaning of this code is not documented in the data we received. 43.1\% are within a municipality: 38.0\% within the 50 municipalities, left out, and 5.1\% within other municipalities, kept in the class with two random ends, of which they are more than a third. The remaining 39.2\% are between municipalities with known ends, on 7,711 pairs: 28\% of these trips keep their relation (476 pairs with agents after rounding) and 72\% have a known relation that is discarded. The relations kept are only 6\% of the pairs, because trips concentrate on few pairs. They are short: mean 13~km, and 87\% lie below 20~km on a Monday. Saturday is similar.

\begin{figure}[htbp]
\centering
\includegraphics[width=0.9\textwidth]{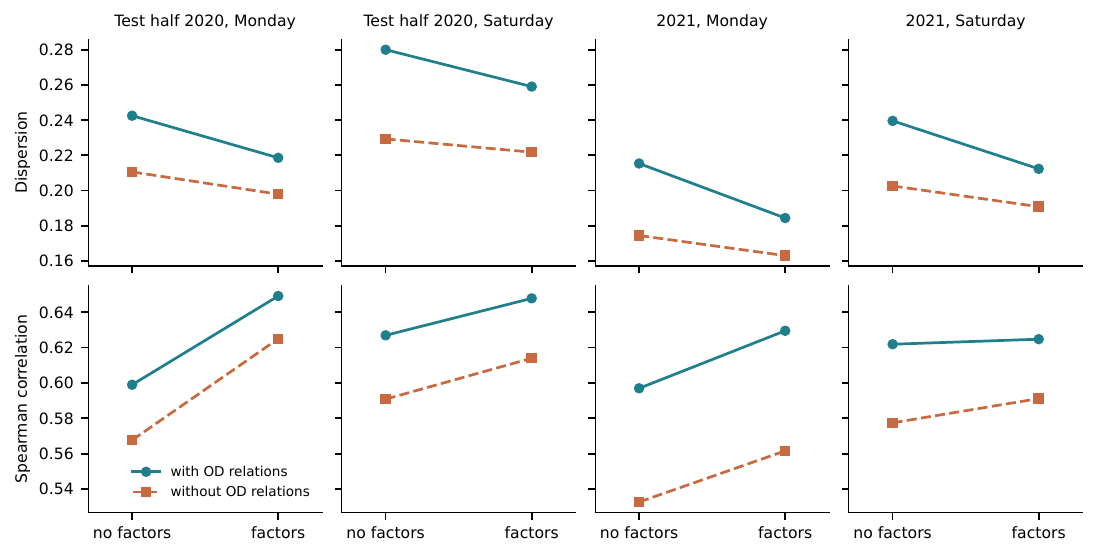}
\caption{\textbf{OD relations and province factors, $2\times 2$ design.} Dispersion (top) and Spearman correlation (bottom) of daily volumes in the four simulations. Parallel lines mean no interaction. Dispersion excludes detectors with no simulated traffic.}
\label{fig:factorial}
\end{figure}

\paragraph{Relations and province factors.} Figure~\ref{fig:factorial} shows the complete design. Table~\ref{tab:od} gives the differences. The interaction in dispersion is negative in all four comparisons ($[-0.022, -0.000]$, $[-0.031, -0.008]$, $[-0.024, -0.004]$, $[-0.025, -0.006]$), so the factors reduce the spread more with relations. In the simulation without relations the regional pattern acts only through random ends.

\paragraph{OD pairings.} Table~\ref{tab:exp_pairing} gives the results. Origins, destinations and the total are identical to the real arm in every interval, and null D keeps the distance classes exactly. Null M raises the mean trip to 66--68~km at 50 municipalities (real 13~km) and 84--86~km at 347 (real 12~km). Null D moves 51\% of agents at 50 municipalities and 67\% at 347; more swaps add little, because the constraints leave many pairs no alternative. The six runs of null M at 347 municipalities (about 88,000 sources) ended with a segmentation fault in an earlier configuration, and these runs were not repeated. The probable cause, not confirmed for these runs, is a defect in a diagnostic output of the simulator that reads an uninitialised value for sources without a location; other runs with many sources failed there, and we corrected this for the tests in the paragraph on routing and trip length below. The 347 arm has a larger dispersion than the 50 arm (0.33--0.47 against 0.19--0.26), so it is used only for this comparison: nearly all its demand is assigned to specific pairs, and only 5\% of its agents on a Monday have a random end, against 72\% in the model, so some detectors away from those pairs have little traffic. The twin test takes a real run as the truth and asks whether a real run with another seed beats the nulls; it has no model error, so it overstates the power. Spearman correlations with GPS link counts of Forl\`i--Cesena differ from the real arm by at most 0.008 at 50 municipalities and 0.003 at 347.

\begin{table}[htbp]
\centering
\caption{OD pairings: real against null M and D, at 50 or 347 municipalities. Differences: real minus mean of three rearrangements; 95\% paired bootstrap over detectors (2,000 resamples, seed 20261061). Sens.: median $|\log_{10}|$(real/null) of daily volume at detectors. Twin: difference in dispersion with a real run as the truth.}
\label{tab:exp_pairing}
\resizebox{\textwidth}{!}{\input{tables/od_pairing}}
\end{table}

\paragraph{Route choice on the final model.} The selected form stands: no other has an interval above zero on both days (Table~\ref{tab:exp_routing}). No other cost weight replaces $(0.5,0.25)$ on the calibration detectors (Table~\ref{tab:exp_alpha}). Table~\ref{tab:exp_separation} separates speeds and usage weights at detectors.

\begin{table}[htbp]
\centering
\caption{Route-choice variants on the final model: Spearman correlation with GPS counts in Forl\`i--Cesena, and 95\% interval of the difference from the selected variant, paired over links.}
\label{tab:exp_routing}
\resizebox{\textwidth}{!}{\input{tables/experiments_routing}}
\end{table}

\begin{table}[htbp]
\centering
\caption{Speeds and usage weights at detectors outside Forl\`i--Cesena, final model: Spearman correlation and 95\% paired intervals. V0: shortest distance, default speeds. V1: shortest distance, GPS speeds. D1: usage cost and class weights, default speeds. V6: usage cost and class weights, GPS speeds.}
\label{tab:exp_separation}
\resizebox{\textwidth}{!}{\input{tables/experiments_separation}}
\end{table}

\begin{table}[htbp]
\centering
\caption{Cost weights on the calibration detectors, final model (selected: dispersion 0.185, Spearman 0.736). Weights: $100\alpha$ and $100\alpha_s$. Interval of the change in dispersion from the selected weights, 95\% paired.}
\label{tab:exp_alpha}
\input{tables/experiments_alpha}
\end{table}

\paragraph{Scale $k$ with province factors.} The selected values remain the minima of the grid (ratios 1.01 and 1.04). The neighbours cannot be told apart on the loss: $k=1.4$ $[-0.001, +0.004]$ and $k=1.6$ $[-0.001, +0.003]$ on weekdays, $k=1.2$ $[-0.002, +0.004]$ and $k=1.4$ $[-0.002, +0.004]$ on weekends.

\paragraph{How many detectors.} With 200 random subsets of the calibration half, 25 detectors choose the selected $k$ or a neighbour in 72\% of cases on weekdays and 80\% on weekends, and 50 detectors in 96\% and 99\%; the mean loss on the test half changes by at most 0.001 between 25 and 50 detectors. Province factors from 25 calibration detectors give a larger dispersion than those from 100 in 9 of 12 comparisons over three subsets, and those from 50 in 5 of 12. The values for a single province vary strongly between subsets: Modena from 0.50 to 1.75 and Piacenza from 1.44 to 2.50, against 1.19 and 1.74 with all 100 detectors. Table~\ref{tab:exp_curve} gives the correction with subsets of detectors. It uses a fixed 20\% of detectors held out on single days, so it is not comparable in size with the leave-one-out test.

\begin{table}[htbp]
\centering
\caption{Correction with subsets of detectors, depth 3, five repeats, on a fixed 20\% of complete detectors held out. Hourly error of the held-out detectors.}
\label{tab:exp_curve}
\input{tables/experiments_curve}
\end{table}

\paragraph{More relations with province factors.} $N=100$, with a multiplier of 1.15 on both day types, does not beat the model on the calibration half: dispersion 0.200 against 0.185 ($[-0.002, 0.030]$), Spearman 0.781 against 0.736 ($[0.004, 0.096]$); the rule requires lower dispersion. The 2021 detectors were not used. The direction of travel does not meet its rule on the test half (median signal 0.046 against 0.054, $[-0.051, 0.048]$).

\paragraph{2021 without calibration detectors.} The 141 detectors of 2021 include 60 of the calibration half. All detectors, and in brackets the 81 others: volume ratio 1.10 (1.11, interval [0.99, 1.24]), Spearman correlation 0.66 (0.65, [0.50, 0.76]), median $|\log_{10}|$ ratio 0.142 (0.146, [0.124, 0.200]), within a factor of two 84\% (81\%, [73, 90]).

\paragraph{Variability between runs.} Between four seeds, the ranges are at most 0.002 for the volume ratio, 0.002 for the Spearman correlation, 0.006 for dispersion and 0.011 for the median error. These ranges may understate the variation between runs, because choices made while agents move use a random generator that does not change with the seed.

\paragraph{What explains the error at single roads.} This diagnostic followed a protocol in the code repository, written before it was computed, and uses no simulations. The outcome is $\log_{10}$(simulated/observed) of daily volume and its absolute value, raw simulation. Table~\ref{tab:road_error} gives the out-of-fold $R^2$ lost when a group of predictors is removed from an ordinary least squares model (5 folds, 100 repeats), with permutation $p$ values Holm-adjusted within outcome. By its rule, a group is relevant if on the Monday it loses at least 0.05 with adjusted $p<0.05$, and at least 0.02 in 2021. Only road class meets it, on the signed error: primary roads are overestimated (median $+0.051$, 94 detectors) and secondary roads underestimated ($-0.036$, 53), while trunk roads have an unstable value ($-0.237$ with 12 detectors, $-0.044$ with 6 in 2021). This diagnostic uses the 164 detectors with weekday data on the Monday, whereas Fig.~\ref{fig:distributions} of the main text uses the 152 with complete days of the leave-one-out test. The full model has an out-of-fold $R^2$ of 0.09 on the Monday, 0.12 on the Saturday and $-0.10$ in 2021, and $-0.33$ on the Monday when one province is left out in turn; a random forest gives 0.04 out of bag on the Monday. No group predicts the absolute error under the rule. The measure of omitted roads counts their length within 1~km, and does not show whether a given road takes traffic from the detector's road. 2021 is not a replication, since detectors and demand are the same.

\begin{table}[htbp]
\centering
\caption{Out-of-fold $R^2$ lost when a group of predictors is removed, signed error, with Holm-adjusted permutation $p$. Omitted: OpenStreetMap roads farther than 30~m from any model link, within 1~km. Parallel: model links within 1~km with a similar bearing. Activity: population and OpenStreetMap points of activity within 1 and 5~km.}
\label{tab:road_error}
\begin{tabular}{@{}lrrrr@{}}
\toprule
Group & Monday (164) & $p$ & 2021 (122) & $p$ \\
\midrule
Road class & 0.194 & 0.007 & 0.090 & 0.042 \\
Province & 0.067 & 0.066 & $-0.055$ & 0.78 \\
Length and centrality & $-0.020$ & 1.00 & $-0.013$ & 0.78 \\
Parallel roads & $-0.025$ & 1.00 & $-0.023$ & 0.78 \\
Omitted roads & $-0.007$ & 0.70 & $-0.059$ & 0.78 \\
Activity & 0.007 & 1.00 & $-0.008$ & 0.48 \\
Distance from the regional boundary & $-0.013$ & 1.00 & $-0.007$ & 0.78 \\
\bottomrule
\end{tabular}
\end{table}

\paragraph{Routing and trip length.} On a Monday, trips between municipalities average 12.0~km in a straight line between centres (median 10.4~km, 96.5\% below 25~km). The simulation gives about 40~km per agent: 42--45~km for agents with a random end and 27~km for agents with a known relation, estimated from hourly positions. These further tests followed a protocol in the code repository, written before their runs, and are exploratory. They used the detectors of the calibration half and half of the test half, by province, in 2020 and 2021; the other half was set aside and not used, and the model was not changed. With 347 municipalities, the share of detectors more than a factor of two below, after removing the level of each arm, is 17.9\% on average in 2020 and 6.5\% in the model. Four route types per relation give 16.8\%, and a probability of 0.10 or 0.30 of leaving the best path at each node gives 17.2\% and 11.9\%, so no arm reaches the threshold of 10\%. Four route types raise the Spearman correlation in 2021 to 0.70, against 0.66 with one route and 0.62 in the model. With the model of 50 municipalities, shorter walks of 900--1,800~s and $k$ selected again at 2.5 on both day types, with province factors estimated again, give a dispersion of 0.272 on the test detectors against 0.222 in the model ($[0.017, 0.091]$ and $[0.021, 0.073]$ on the two days), and a Spearman correlation of 0.73 against 0.72, and of 0.66 against 0.62 in 2021, with intervals that include zero. Walks of exactly 900~s, with or without a radius of 10~km for known relations, and with the factors of the model, give dispersions of 0.28--0.30 and Spearman correlations of 0.75--0.76. By the rule fixed before the runs, these results do not replace the model. The tests with four route types and those on trip length used a simulator with the diagnostic output corrected; on the final Monday it reproduces the link flows byte for byte.

\paragraph{Route weights by link.} Following a protocol written before the runs, 398 links of Forl\`i--Cesena with at least 50 probes in February 2020 received a weight from their own GPS count, on the scale of the class weights, with everything else unchanged, for one Monday and one Saturday. The simulation moves towards GPS on those links (Spearman from 0.66 to 0.77, in sample). At the detectors of the province, which did not enter the weights, the Spearman differences have intervals that include zero, and the dispersion falls on the Monday of February 2020 (0.228 to 0.176, $[-0.091, -0.016]$, 18 detectors) but not on the Saturday (0.314 to 0.326, $[-0.054, +0.059]$, 19) or in 2021 (Monday 0.177 to 0.190, $[-0.032, +0.053]$, 7; Saturday 0.224 to 0.200, $[-0.072, +0.035]$, 11). The rule is not met, and the model keeps the class weights. With so few detectors the test is inconclusive.

\section{Value of the OD relations}
\label{sec:od_value}

To measure the value of the OD relations, we ran one simulation per day type without them. We kept the number of agents per interval and $k$, but drew all origins and destinations from population. The province factors act through sources with a location, so this simulation loses them too. Under a rule fixed before the runs, the relations are supported if, on the test half of 2020 or on the detectors of 2021, the 95\% interval of the Spearman difference is above zero or that of the dispersion difference is below zero. We also compared with GPS link counts, and we repeated the runs with further seeds to measure the variability between them. We also tried more relations: on the calibration detectors only, we let the relations among the 100 most active municipalities keep both ends, with a new multiplier of $k$ (Section~\ref{sec:experiments}).

Further experiments followed protocols written before their runs (Section~\ref{sec:experiments}). They are exploratory, because the data had already been seen, and they do not change the model. To separate relations from factors, we ran the simulation without relations with node scores multiplied by the square root of the province factor, which gives a trip the geometric mean of the factors of its ends, and we completed a $2\times 2$ design. To isolate the pairings, we rearranged the destinations of agents with located relations within each 15-min interval, three times per case, at 50 and 347 municipalities. Origins, destinations and the total are unchanged. In null M the rearrangement is random, while in null D each agent stays in the distance class of its real trip (0--5, 5--10, 10--20, 20--50, above 50~km). Real pairs are supported if they beat null D in at least two of four comparisons on the same metric.

\begin{figure}[htbp]
\centering
\includegraphics[width=\textwidth]{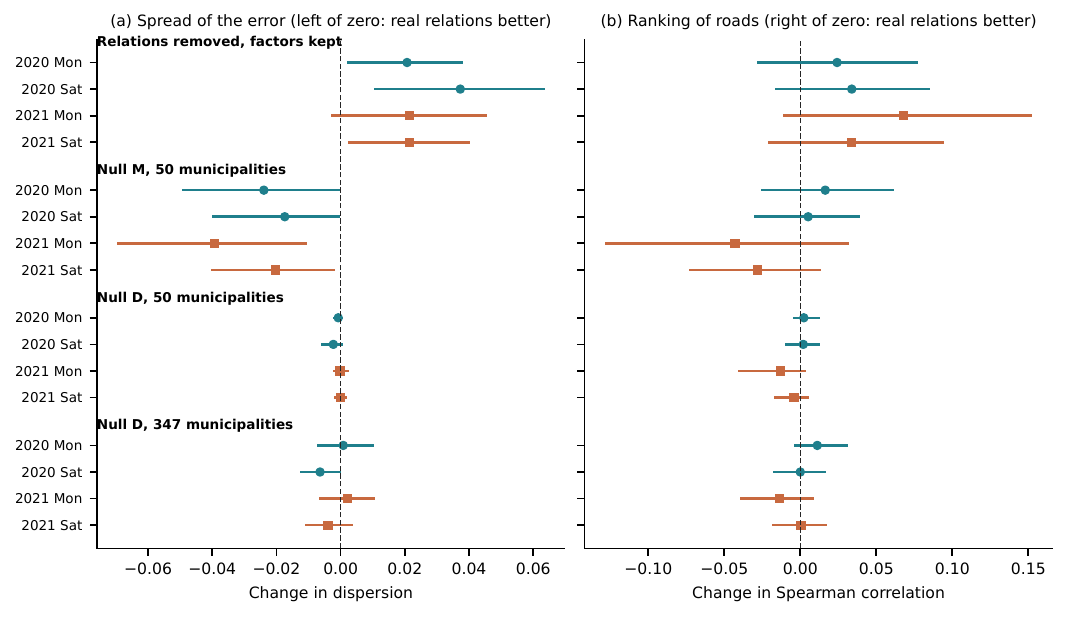}
\caption{\textbf{What the OD relations add.} Each row compares the model with real relations with a version without them: the marker is the estimated change and the bar its 95\% bootstrap interval, so a bar crossing the dashed line means that the data cannot establish a change. Blue circles: test half, 2020; orange squares: all detectors, 2021; Monday and Saturday simulations. \textbf{Relations removed, factors kept}: no located relations, with the regional pattern of the factors kept in the random ends. \textbf{Null M}: destinations rearranged at random within each 15-min interval, so that origins and destinations by municipality are exact and distances are lost. \textbf{Null D}: the same, with each agent kept in the distance class of its real trip. The null rows are the mean of three rearrangements. \textbf{(a)} Spread of the error between roads. \textbf{(b)} Ranking of roads. Real relations tend to have a lower spread than random pairs, and cannot be told apart from pairs with the same distances.}
\label{fig:od}
\end{figure}

We now ask what the OD relations themselves add. Only part of the table can answer: on a Monday, 28\% of the trips between municipalities with known ends keep their relation, 11\% of the table (Section~\ref{sec:experiments}). Removing the relations from the final model takes away the regional pattern of the factors as well. In that comparison the model with relations has the higher Spearman correlation on the test half (0.08 [0.02, 0.15] on the Monday and 0.06 [0.00, 0.12] on the Saturday) and on the Monday of 2021 (0.10 [0.01, 0.20]), which meets the rule fixed before the runs, while the Saturday interval of 2021 reaches zero and dispersion does not improve (Section~\ref{sec:additional}). When the same pattern is kept in the simulation without relations, the Spearman differences stay positive with intervals that include zero, and the model with relations has a larger spread of errors in all four comparisons, with intervals that exclude zero in three (Fig.~\ref{fig:od}, top; Table~\ref{tab:od}). The factors reduce the spread more when relations are present (Section~\ref{sec:experiments}). Chance between runs is small compared with these differences.

Removing relations also changes how many trips start and end in each place. To isolate the pairings, we kept these numbers exact in each 15-min interval and rearranged only who goes where. Random pairs make the trips far too long, and they raise dispersion, with intervals that exclude zero in all four comparisons, two of them narrowly (Fig.~\ref{fig:od}, middle); agreement in ranking does not change measurably. When distances are kept, real and rearranged pairs can hardly be told apart (Fig.~\ref{fig:od}, bottom). At 50 municipalities no dispersion interval excludes zero and at 347 one does, so the rule fixed before the runs (two of four) is not met. Agreement with GPS link counts does not change either. These tests have limited power: rearranging with distances kept changes the volume at the median detector by only 1--3\%, and a test with a simulated run as the truth shows that the expected effect at 50 municipalities is about as small as the intervals (Section~\ref{sec:experiments}). Extending the relations to 100 municipalities with province factors does not beat the model under its rule (Section~\ref{sec:experiments}).

The simulation without relations keeps the agents per interval, the daily rhythm and the population, so it is not a random traffic pattern. In an analysis made after the results above, without detector 153, the larger spread with relations comes mainly from the low tail in 2020: 10--14\% of detectors are more than a factor of two below with relations, against 5\% without. With a spread measured from the quartiles no interval excludes zero, and the Spearman correlation is higher with relations in all four comparisons, with intervals that include zero. The median error differs in one comparison of four, in favour of the simulation without relations. So these detectors do not show that random ends are better than relations.

Null D with its five distance classes can still change trip length within a class, and it could not be read as a test of pairings alone. We therefore registered a further test, null D2, with a measured power rule: each radius is classified only if the power of the rule reaches 80\%, and otherwise it is inconclusive whatever the interval. The protocol asked for distance bins of 2~km. These failed its check on the share of agents moved, so before any D2 run an amendment set them to 4~km; the width was chosen knowing that share (51--52\% of agents moved, mean distance within 0.3\%), and it clears the threshold narrowly. Each arm uses two real runs and three rearrangements at 347 municipalities, and pools 368 rows on 168 detectors. The estimand is the slope $b$ of the observed log difference from the null on the simulated log difference between real and null, estimated with a cross-fitted instrumental variable, with $b=1$ expected if real pairs act as simulated and $b=0$ if they add nothing. Table~\ref{tab:d2} gives the results. Both radii are inconclusive, with a tau/sigma ratio of 0.01 and 0.06 against the 0.3 needed for 80\% power. At 25~km the estimate has no meaning, because the signal is near zero. At 10~km, the primary radius of the protocol, tested only here, the interval excludes both 0 and 1 and the volume is 40--45\% too low, with $k$ not retuned; under the simulated power analysis an interval excludes zero in 6\% of cases if $b=0$ and in 14\% if $b=1$. So this estimate cannot be read as the value of the pairings. The arm also differs from the reported model, as described in Section~\ref{sec:experiments}. The value of specific origin--destination pairings could not be determined from link-count observations with the statistical power available in this study; the outcome is inconclusive, and it is not evidence that the pairings are unnecessary.

\begin{table}[htbp]
\centering
\caption{Null D2 test, 347 municipalities, 168 detectors, 368 rows. Interval: 95\% bootstrap over detectors. $\tau/\sigma$: measured signal over residual spread. Power: of the registered rule, at the measured $\tau/\sigma$. Exploratory, because the data had been seen.}
\label{tab:d2}
\begin{tabular}{@{}lrrrrl@{}}
\toprule
Radius & $b$ & Interval & $\tau/\sigma$ & Power & Classification \\
\midrule
25~km, as in the model & not interpretable & $[-233, 195]$ & 0.011 & 0.00 & inconclusive \\
10~km & 4.58 & $[2.39, 7.37]$ & 0.063 & 0.11 & inconclusive \\
\bottomrule
\end{tabular}
\end{table}

\paragraph{Winter 2023.} Following a protocol written before any simulated value was compared with a 2023 count, we compared the model with mean detector counts of winter 2023, outside the calibration half. This is not a replication, because the simulations are the same and the detectors overlap with the earlier sets, and the dates and holidays of the period are not documented in the data we received. Set A (all detectors outside the calibration half) gives a Spearman correlation of 0.74 [0.65, 0.81] on the Monday (122 detectors) and 0.73 [0.63, 0.80] on the Saturday (140); set B (those also in the 2020 or 2021 sets) gives 0.70 [0.57, 0.79] (89) and 0.67 [0.53, 0.78] (99); set C (the others) gives 0.85 [0.72, 0.91] (33) and 0.85 [0.72, 0.91] (41). Most of the apparent gain comes from which detectors are in the set: on common detectors the year changes little (Monday test half 0.63 in 2020 and 0.63 in 2023; Saturday test half 0.62 and 0.62; Monday 2021 0.64 and 0.66; Saturday 2021 0.64 and 0.69). The means over many days also lower the noise, and set C is not untouched, because earlier rounds scanned $k$ on means of 2022--24. The volume is 39\% above the 2023 counts on the Monday and 12\% above on the Saturday; we report this as a description only. Null D2 on set A gives the same classification at both radii.

\begin{table}[htbp]
\centering
\caption{OD relations and province factors in a complete $2\times 2$ design: difference from removing the relations with the factors kept in the random ends, and from adding the factors to the simulation without relations. Positive Spearman differences and negative dispersion differences mean a gain from adding the source. Intervals: 95\% paired bootstrap over detectors. Exploratory.}
\label{tab:od}
\input{tables/od_factorial}
\end{table}

\end{document}

%% file: tables/routing_variants.tex
\begin{table}[htbp]\centering
\caption{Route-choice variants simulated for 10 February 2020. Calibration: Spearman correlation between simulated daily link volumes and GPS counts on Forl\`i--Cesena links. Detectors: Spearman correlation with the daily volumes of the detectors outside Forl\`i--Cesena complete on that day (not used for selection). Main roads: share of simulated vehicle-kilometres on motorway, trunk and primary links.}
\label{tab:routing}
\resizebox{\textwidth}{!}{\begin{tabular}{@{}lrrrrr@{}}\toprule
Variant & GPS links & Calibration $\rho_s$ & Detectors & Detector $\rho_s$ & Main roads \\\midrule
Shortest distance, default speeds & 814 & 0.474 & 53 & 0.198 & 0.39 \\
Shortest distance, GPS speeds & 814 & 0.478 & 53 & 0.190 & 0.38 \\
Travel time, GPS speeds & 814 & 0.487 & 53 & 0.167 & 0.38 \\
Mixed distance/time, GPS speeds & 814 & 0.484 & 53 & 0.175 & 0.37 \\
Travel time + class transition weights & 814 & 0.537 & 53 & 0.109 & 0.47 \\
Travel time + damped class weights & 814 & 0.532 & 53 & 0.105 & 0.42 \\
Usage cost + class transition weights (selected) & 814 & 0.542 & 53 & 0.176 & 0.48 \\
\bottomrule\end{tabular}}\end{table}

%% file: tables/routing_volumes.tex
\begin{table}[htbp]\centering
\caption{Daily volumes of the simulation before detector correction at the detectors outside Forl\`i--Cesena: Spearman correlation with the observed volumes for shortest-distance and GPS-informed route choice, paired bootstrap difference with 95\% interval (5,000 resamples of detectors), and median absolute $\log_{10}$ ratio of simulated to observed volume.}
\label{tab:routing_volumes}
\resizebox{\textwidth}{!}{\begin{tabular}{@{}lrrrlrr@{}}\toprule
Campaign & Detectors & $\rho_s$ distance & $\rho_s$ GPS & Difference & $|\log_{10}|$ distance & $|\log_{10}|$ GPS \\\midrule
2020 & 136 & 0.64 & 0.72 & 0.08 [0.02, 0.15] & 0.14 & 0.13 \\
2021 & 128 & 0.54 & 0.64 & 0.10 [0.03, 0.18] & 0.15 & 0.14 \\
\bottomrule\end{tabular}}\end{table}

%% file: tables/routing_shape.tex
\begin{table}[htbp]\centering
\caption{Median hourly-share MAE (pp) of the simulation before detector correction on GPS links of Forl\`i--Cesena, with shortest-distance and GPS-informed route choice.}
\label{tab:routing_shape}
\begin{tabular}{@{}llrr@{}}\toprule
Campaign & Route choice & Links & MAE (pp) \\\midrule
2020 & shortest distance & 442 & 1.06 \\
2020 & GPS-informed & 442 & 1.08 \\
2021 & shortest distance & 439 & 1.20 \\
2021 & GPS-informed & 439 & 1.21 \\
\bottomrule\end{tabular}\end{table}

%% file: tables/loo.tex
\begin{table}[htbp]\centering
\caption{Leave-one-detector-out test. Errors are $\sum|\tilde u-c|/\sum c$ in percent of the observed volume, pooled over detectors or as the median over detectors; reached: share of left-out detectors whose link is modified by the spreading from other detectors; improved/worse: share with lower/higher error than the simulation before correction; $\rho_s$: Spearman correlation of daily volumes with the observed ones before correction and with the detector left-out.}
\label{tab:loo}
\resizebox{\textwidth}{!}{\begin{tabular}{@{}lrrrrrrrrrrr@{}}\toprule
Year & Detectors & Detector-days & Pooled raw & Pooled LOO & Median raw & Median LOO & Reached (\%) & Improved (\%) & Worse (\%) & $\rho_s$ raw & $\rho_s$ LOO \\\midrule
2020 & 152 & 614 & 42.2 & 39.8 & 35.9 & 33.3 & 86 & 60 & 26 & 0.72 & 0.72 \\
2021 & 141 & 572 & 47.4 & 41.4 & 39.4 & 34.8 & 85 & 62 & 23 & 0.66 & 0.69 \\
\bottomrule\end{tabular}}\end{table}

%% file: tables/depth.tex
\begin{table}[htbp]\centering
\caption{Reduction of the pooled error (percentage points of observed volume) on left-out detectors, 5-fold cross-validation over the 152 detectors of 2020 complete on each day, with bootstrap 95\% intervals over detectors; difference: paired with the selected depth 3, the smallest within one standard error (1.5 pp) of the best depth. Raw error: 42.2\%.}
\label{tab:depth}
\resizebox{\textwidth}{!}{\begin{tabular}{@{}rrlrl@{}}\toprule
Depth (links) & Error reduction (pp) & 95\% interval & Difference (pp) & 95\% interval \\\midrule
1 & 0.9 & [-0.0, 2.2] & -1.4 & [-3.3, +0.2] \\
2 & 1.2 & [-0.3, 2.9] & -1.1 & [-2.2, -0.3] \\
3 & 2.3 & [0.4, 4.7] & -- & -- \\
4 & 2.5 & [0.1, 5.1] & +0.1 & [-0.6, +0.8] \\
5 & 3.0 & [0.5, 5.9] & +0.7 & [-0.2, +1.5] \\
6 & 3.2 & [0.5, 6.2] & +0.8 & [-0.2, +1.8] \\
\bottomrule\end{tabular}}\end{table}

%% file: tables/probe_strata.tex
\begin{table}[htbp]\centering
\caption{Hourly-share MAE (pp) on GPS links of Forl\`i--Cesena, by number of GPS probes per link and campaign: reconstruction, mean detector profile and hourly correlation of the reconstruction (medians).}
\label{tab:probes}
\begin{tabular}{@{}llrrrr@{}}\toprule
Campaign & Probes & Links & Reconstruction & Mean detector & Correlation \\\midrule
2020 & 1-49 & 56 & 4.49 & 4.49 & 0.46 \\
2020 & 50-199 & 79 & 1.86 & 1.82 & 0.75 \\
2020 & 200+ & 307 & 0.85 & 0.81 & 0.93 \\
2021 & 1-49 & 64 & 3.86 & 3.89 & 0.35 \\
2021 & 50-199 & 110 & 1.81 & 1.81 & 0.72 \\
2021 & 200+ & 265 & 0.99 & 0.86 & 0.89 \\
\bottomrule\end{tabular}\end{table}

%% file: tables/od_decomposition.tex
\begin{table}[htbp]\centering
\caption{Separating the OD relations from the province factors with existing simulations. Differences in Spearman correlation and dispersion of daily volumes, with 95\% paired bootstrap intervals over detectors (2,000 resamples, seed 20261049). The last block repeats the comparison that also removes the regional pattern of the factors, with a new bootstrap, so its intervals can differ slightly from those of Section S5. Exploratory: added after the results. Dispersion excludes detectors with no simulated traffic.}
\label{tab:od_decomposition}
\small
\begin{tabular}{@{}lllrll@{}}\toprule
Comparison & Detectors & Day & $n$ & Spearman & Dispersion \\\midrule
Relations, no factors & Test half, 2020 & Monday & 82 & +0.031 [-0.024, +0.085] & +0.032 [+0.014, +0.051] \\
Relations, no factors & Test half, 2020 & Saturday & 93 & +0.036 [-0.015, +0.093] & +0.051 [+0.021, +0.080] \\
Relations, no factors & 2021 & Monday & 76 & +0.064 [-0.030, +0.170] & +0.041 [+0.013, +0.070] \\
Relations, no factors & 2021 & Saturday & 120 & +0.044 [-0.010, +0.103] & +0.037 [+0.017, +0.057] \\
Factors, with relations & Test half, 2020 & Monday & 82 & +0.050 [-0.004, +0.115] & -0.024 [-0.041, -0.007] \\
Factors, with relations & Test half, 2020 & Saturday & 93 & +0.021 [-0.027, +0.075] & -0.021 [-0.037, -0.007] \\
Factors, with relations & 2021 & Monday & 76 & +0.033 [-0.059, +0.122] & -0.031 [-0.048, -0.014] \\
Factors, with relations & 2021 & Saturday & 120 & +0.003 [-0.041, +0.049] & -0.027 [-0.042, -0.014] \\
Relations and factors & Test half, 2020 & Monday & 82 & +0.080 [+0.015, +0.147] & +0.008 [-0.011, +0.027] \\
Relations and factors & Test half, 2020 & Saturday & 93 & +0.058 [+0.001, +0.119] & +0.029 [-0.001, +0.059] \\
Relations and factors & 2021 & Monday & 76 & +0.102 [+0.010, +0.206] & +0.009 [-0.013, +0.031] \\
Relations and factors & 2021 & Saturday & 120 & +0.052 [-0.003, +0.115] & +0.010 [-0.007, +0.027] \\
\bottomrule\end{tabular}
\end{table}

%% file: tables/od_pairing.tex
\begin{tabular}{@{}lllrrrlrrlrl@{}}\toprule
Null & Detectors & Day & $n$ & \multicolumn{3}{c}{Dispersion: real, null, difference} & \multicolumn{3}{c}{Spearman: real, null, difference} & Sens. & Twin, disp. \\\midrule
M, 50 & Test half, 2020 & Monday & 82 & 0.217 & 0.241 & [-0.049, -0.000] & 0.670 & 0.654 & [-0.025, +0.061] & 0.022 & [-0.112, -0.050] \\
M, 50 & Test half, 2020 & Saturday & 93 & 0.264 & 0.281 & [-0.040, -0.000] & 0.663 & 0.658 & [-0.030, +0.038] & 0.021 & [-0.103, -0.046] \\
M, 50 & 2021 & Monday & 76 & 0.186 & 0.225 & [-0.069, -0.011] & 0.640 & 0.684 & [-0.128, +0.032] & 0.043 & [-0.154, -0.085] \\
M, 50 & 2021 & Saturday & 120 & 0.214 & 0.234 & [-0.040, -0.002] & 0.638 & 0.666 & [-0.072, +0.012] & 0.022 & [-0.124, -0.065] \\
D, 50 & Test half, 2020 & Monday & 82 & 0.217 & 0.218 & [-0.002, +0.000] & 0.670 & 0.668 & [-0.004, +0.012] & 0.005 & [-0.003, +0.001] \\
D, 50 & Test half, 2020 & Saturday & 93 & 0.264 & 0.266 & [-0.006, +0.000] & 0.663 & 0.661 & [-0.009, +0.012] & 0.007 & [-0.004, +0.001] \\
D, 50 & 2021 & Monday & 76 & 0.186 & 0.186 & [-0.002, +0.002] & 0.640 & 0.654 & [-0.040, +0.003] & 0.005 & [-0.005, -0.001] \\
D, 50 & 2021 & Saturday & 120 & 0.214 & 0.214 & [-0.002, +0.002] & 0.638 & 0.642 & [-0.016, +0.005] & 0.006 & [-0.004, -0.001] \\
D, 347 & Test half, 2020 & Monday & 82 & 0.428 & 0.427 & [-0.007, +0.010] & 0.652 & 0.641 & [-0.003, +0.030] & 0.014 & [-0.044, -0.014] \\
D, 347 & Test half, 2020 & Saturday & 93 & 0.469 & 0.475 & [-0.012, -0.000] & 0.643 & 0.643 & [-0.017, +0.016] & 0.013 & [-0.036, -0.008] \\
D, 347 & 2021 & Monday & 76 & 0.327 & 0.325 & [-0.006, +0.010] & 0.689 & 0.703 & [-0.039, +0.008] & 0.011 & [-0.022, -0.004] \\
D, 347 & 2021 & Saturday & 120 & 0.414 & 0.418 & [-0.011, +0.003] & 0.656 & 0.656 & [-0.018, +0.017] & 0.012 & [-0.024, -0.012] \\
\bottomrule\end{tabular}

%% file: tables/experiments_routing.tex
\begin{tabular}{@{}lrlrl@{}}\toprule
Variant & Monday & minus selected & Saturday & minus selected \\\midrule
Shortest distance, default speeds & 0.650 & [-0.074, +0.002] & 0.654 & [-0.088, -0.022] \\
Shortest distance, GPS speeds & 0.663 & [-0.062, +0.012] & 0.661 & [-0.081, -0.015] \\
Travel time, GPS speeds & 0.668 & [-0.044, +0.009] & 0.682 & [-0.050, -0.005] \\
Mixed distance/time, GPS speeds & 0.678 & [-0.040, +0.024] & 0.685 & [-0.052, +0.005] \\
Travel time + class weights & 0.690 & [-0.010, +0.015] & 0.715 & [-0.005, +0.016] \\
Travel time + damped class weights & 0.680 & [-0.024, +0.010] & 0.702 & [-0.022, +0.007] \\
Usage cost + class weights, default speeds & 0.676 & [-0.016, -0.005] & 0.701 & [-0.014, -0.004] \\
Usage cost + class weights (selected) & 0.687 & -- & 0.710 & -- \\
\bottomrule\end{tabular}

%% file: tables/experiments_separation.tex
\begin{tabular}{@{}llrrrrrlll@{}}\toprule
Detectors & Day & $n$ & V0 & V1 & D1 & V6 & V1$-$V0 & D1$-$V0 & V6$-$D1 \\\midrule
Test half, 2020 & Monday & 73 & 0.60 & 0.64 & 0.66 & 0.68 & [+0.01, +0.07] & [-0.04, +0.16] & [-0.01, +0.05] \\
Test half, 2020 & Saturday & 84 & 0.56 & 0.58 & 0.66 & 0.68 & [+0.00, +0.04] & [+0.01, +0.19] & [-0.01, +0.05] \\
2021 & Monday & 69 & 0.62 & 0.63 & 0.63 & 0.65 & [-0.01, +0.04] & [-0.10, +0.13] & [-0.01, +0.05] \\
2021 & Saturday & 109 & 0.48 & 0.50 & 0.62 & 0.61 & [-0.01, +0.04] & [+0.03, +0.24] & [-0.03, +0.02] \\
\bottomrule\end{tabular}

%% file: tables/experiments_alpha.tex
\begin{tabular}{@{}lrrl@{}}\toprule
Weights & Dispersion & Spearman & Change in dispersion \\\midrule
w100, s000 & 0.190 & 0.729 & [-0.004, +0.013] \\
w075, s025 & 0.188 & 0.735 & [-0.003, +0.008] \\
w050, s050 & 0.188 & 0.734 & [-0.002, +0.008] \\
w025, s075 & 0.184 & 0.735 & [-0.007, +0.006] \\
w000, s100 & 0.184 & 0.734 & [-0.013, +0.012] \\
w025, s025 & 0.189 & 0.729 & [-0.002, +0.010] \\
w050, s000 & 0.190 & 0.726 & [+0.001, +0.009] \\
\bottomrule\end{tabular}

%% file: tables/experiments_curve.tex
\begin{tabular}{@{}lrrrrrl@{}}\toprule
Day & Share (\%) & Detectors & Raw (\%) & Corrected (\%) & Gain (pp) & Range over repeats \\\midrule
2020-02-10 & 25 & 46 & 46.5 & 45.1 & +1.4 & [-0.1, +3.7] \\
2020-02-10 & 50 & 91 & 46.5 & 44.0 & +2.5 & [+0.3, +3.6] \\
2020-02-10 & 75 & 136 & 46.5 & 43.6 & +2.9 & [+0.6, +4.0] \\
2020-02-10 & 100 & 182 & 46.5 & 43.4 & +3.1 & [+2.9, +3.7] \\
2021-06-28 & 25 & 41 & 42.6 & 42.7 & -0.1 & [-1.6, +1.0] \\
2021-06-28 & 50 & 82 & 42.6 & 41.8 & +0.8 & [-1.4, +3.7] \\
2021-06-28 & 75 & 123 & 42.6 & 41.5 & +1.1 & [-1.5, +3.8] \\
2021-06-28 & 100 & 164 & 42.6 & 41.3 & +1.4 & [-1.3, +4.3] \\
\bottomrule\end{tabular}

%% file: tables/od_factorial.tex
\begin{tabular}{@{}llccc@{}}\toprule
 & & \multicolumn{2}{c}{OD relations, with factors} & Factors, no relations \\
Detectors & Day & Spearman & Dispersion & Dispersion \\\midrule
Test half, 2020 & Monday & +0.024 [-0.027, +0.077] & +0.021 [+0.002, +0.038] & -0.013 [-0.026, +0.001] \\
Test half, 2020 & Saturday & +0.034 [-0.016, +0.085] & +0.037 [+0.011, +0.063] & -0.008 [-0.021, +0.006] \\
2021 & Monday & +0.068 [-0.011, +0.152] & +0.021 [-0.002, +0.045] & -0.011 [-0.026, +0.003] \\
2021 & Saturday & +0.034 [-0.021, +0.094] & +0.022 [+0.003, +0.040] & -0.012 [-0.024, -0.000] \\
\bottomrule\end{tabular}